\documentclass{emulateapj}
\usepackage{apjfonts}
\newcommand\cm{{\;\rm cm}}
\newcommand\pcc{{\;\rm cm^{-3}}}
\newcommand\Kel{{\;\rm K}}

\newcommand\yr{{\;\rm yr}}

\newcommand\Msun{{\;\rm\,M_\odot}}
\newcommand\Surf{{\;\rm\,M_\odot\;pc^{-2}}}

\newcommand\kms{{\;\rm km\,s^{-1}}}

\newcommand\SigNG{\Sigma_{\rm NG}}
\newcommand\torb{t_{\rm orb} }
\newcommand\pc{{\;\rm\,pc}}
\newcommand\kpc{{\;\rm kpc}}

\newcommand\simgt{\lower.5ex\hbox{$\; \buildrel > \over \sim \;$}}
\newcommand\simlt{\lower.5ex\hbox{$\; \buildrel < \over \sim \;$}}
\newcommand\xhat{\hat{\mathbf{x}} }
\newcommand\yhat{\hat{\mathbf{y}} }

\newcommand\vel{\mathbf{v}}
\newcommand\ergs{{\rm \;erg\,s^{-1}}}
\newcommand\kb{k_{\rm B}}
\newcommand\vd{\langle{\delta v}_i\rangle}

\shorttitle{SPIRAL SHOCKS WITH THERMAL INSTABILITY}
\shortauthors{KIM, KIM, \& OSTRIKER}
\begin{document}

\title{Galactic Spiral Shocks with Thermal Instability in Vertically
Stratified Galactic Disks}

\author{Chang-Goo Kim\altaffilmark{1}, Woong-Tae Kim\altaffilmark{1},
and Eve C.\ Ostriker\altaffilmark{2}}
\affil{$^1$Department of Physics \& Astronomy, FPRD,
Seoul National University, Seoul 151-742, Republic of Korea}
\affil{$^2$Department of Astronomy, University of Maryland,
College Park, MD 20742, USA}
\email{kimcg@astro.snu.ac.kr, wkim@astro.snu.ac.kr, ostriker@astro.umd.edu}
\slugcomment{Accepted for publication in \apj}

\begin{abstract}
Galactic spiral shocks are dominant morphological features and
believed to be responsible for substructure formation within spiral
arms in disk galaxies. They can also contribute a substantial amount
of kinetic energy to the interstellar gas by tapping the
(differential) rotational motion. We use numerical hydrodynamic
simulations to investigate dynamics and structure of spiral shocks
with thermal instability in vertically stratified galactic disks,
focusing on environmental conditions (of heating and the galactic
potential) similar to the Solar neighborhood. We initially consider
an isothermal disk in vertical hydrostatic equilibrium and let it
evolve subject to interstellar cooling and heating as well as a
stellar spiral potential. Due to thermal instability, a disk with
surface density $\Sigma_0 \geq 6.7\Surf$ rapidly turns to a thin
dense slab near the midplane sandwiched between layers of rarefied
gas. The imposed spiral potential leads to a vertically curved shock
that exhibits strong flapping motions in the plane perpendicular to
the arm.  The overall flow structure at saturation is
comprised of arm, postshock expansion zone, and interarm regions
that occupy typically 10\%, 20\%, and 70\% of the arm-to-arm
distance, in which the gas resides for 15\%, 30\%, and 55\% of the
arm-to-arm crossing time, respectively. The flows are characterized
by transitions from rarefied to dense phases at the shock and from
dense to rarefied phases in the postshock expansion zone, although
gas with too-large postshock-density does not undergo this return
phase transition, instead forming dense condensations. If
self-gravity is omitted, the shock flapping drives random motions in
the gas, but only up to $\sim 2-3 \kms$  in the in-plane direction
and less than $2\kms$ in the vertical direction. Time-averaged shock
profiles show that the spiral arms in stratified disks are broader
and less dense compared to those in unstratified models, and that
the vertical density distribution is overall consistent with local
effective hydrostatic equilibrium. Inclusion of self-gravity
increases the dense gas fraction by a factor $\sim 2$ and raises the
in-plane velocity dispersion to $\sim 5-7 \kms$. When the disks are
massive enough, with $\Sigma_0 \geq 5\Surf$, self-gravity promotes
formation of bound clouds that repeatedly collide with each other in
the arm and break up in the postshock expansion zone.
\end{abstract}

\section{Introduction}

Spiral arms in disk galaxies are regions of ongoing active star formation,
sharply outlined by bright young star complexes.
They usually span the entire optical disks and sometimes extend even to
outer gaseous disks (e.g., \citealt{dic90,boo08,ber09} and references therein).
Such large-scale spiral patterns may be the manifestation of spiral density
waves which propagate with a constant pattern speed through
stellar disks \citep{lin64,lin66}, or
may be transient features driven, for example, by tidal interactions
with companion galaxies (e.g., \citealt{too72,her90,sal00,oh08,dob09}).
Regardless of the origin of spiral features, it is widely accepted  that
the interstellar medium (ISM) is strongly compressed when
it encounters stellar arms, forming narrow dust lanes in optical images
(e.g., \citealt{elm06,she07}).
The densest parts of the shocked layers subsequently undergo gravitational
collapse and produce downstream secondary structures,
such as OB star complexes and
giant \ion{H}{2} regions (e.g., \citealt{baa63,elm83}),
filamentary gaseous spurs (also referred to as ``feathers;''
e.g., \citealt{sco01,ken04,wil04,lav06,gor07,cor08}), and
giant molecular associations or atomic superclouds
(e.g., \citealt{elm83,vog88,ran90,kod09}).

Since shock compression within the arms is the first step
towards star formation in grand-design spiral galaxies, understanding structural and
dynamical evolution of these gas flows is essential
to a host of fundamental problems, such as global star formation rates,
the nature of the Hubble sequence, galaxy evolution, etc.
Since the pioneering work of \citet{rob69} who obtained
one-dimensional stationary shock profiles,
there have been numerous studies of the structure of
galactic spiral shocks under the simplifying assumption that the gas
remains isothermal
(e.g., \citealt{woo75,woo76,lub86,mar98,kim02,gom02,gom04,wad04,
bol06,kim06, dob06}).
In particular, \citet{woo75} and \citet{kim02} showed that the
one-dimensional shock profiles found by \citet{rob69} represent
stable equilibria when the fluid quantities are allowed to vary only in
the direction perpendicular to the arms.  The growth of axisymmetric
self-gravitating modes is limited by postshock expansion
\citep{lub86}.

When the direction parallel to the arm is included in models, on the other hand,
isothermal spiral shocks in two dimensions are prone to various kinds of
instabilities.  \citet{bal88} showed that when self-gravity is
included, spiral arms
are unstable to transient swing instability. When
magnetic fields are included, spiral arms are subject to magneto-Jeans
instability, in which embedded parallel magnetic fields
that exchange angular momentum limit the
stabilizing effect of galaxy rotation, encouraging non-axisymmetric
perturbations to grow into giant clouds and other arm substructures
\citep{elm94,kim02,kim06,she06}.
\citet{wad04} showed that spiral shocks in two-dimensional
thin disk  models are unstable to a vorticity-generating wiggle
instability and develop arm substructures (see also
\citealt{dob06}), although these in-plane
modes appear to be suppressed by embedded
magnetic fields \citep{she06,dob08b} or by vertical shear and mixing
when all three dimensions are included in models \citep{kim06}.

While steady in-plane shock solutions are subject to instability,
shock models that include the vertical degree of freedom do not even
have steady solutions.  Instead, the shock front in vertically stratified disks
moves back and forth relative to the mean
position \citep{kim06}.  These shock ``flapping'' motions arise mainly
because the vertical oscillation period of the gas is, in general,
incommensurate with
the arm-to-arm crossing time, so that the gas streamlines are not closed.
\citet[hereafter Paper I]{kko06} showed that the shock flapping is able to
feed random gas motions on the scale of
disk scale height that persist despite dissipation.
The induced gas velocity dispersions reach a sonic level,
suggesting that spiral shocks may be a considerable source of
the ISM turbulence.  Motions driven by shock flapping motions
destroy any coherent vortical structures that would otherwise
grow near the spiral shocks, suppressing the wiggle instability.
Since gravity is a long-range force and insensitive to small scale density
structure, however, magneto-Jeans instabilities
still grow within the arms in three-dimensional disk models,
in spite of non-steady motions induced by shock flapping
\citep{kim06}.

Phase transitions caused by thermal instability (TI) create a
multiphase ISM, with important consequences for galactic star
formation. In the classical picture of the ISM, TI changes an
otherwise uniform ISM into warm rarefied material and cold dense
clouds in a rough pressure balance (e.g.,
\citealt{fie65,fie69,mee96,hei01,wol03}), while there also exists
significant mass in the thermally-unstable temperature range (e.g.,
\citealt{hei03}). Supernova blast waves create an additional, hot
component that is organized into bubbles or cavities
\citep{cox74,mo77}, although the total mass contained in the hot
phase is much smaller than that in the cold and warm phases (e.g.,
\citealt{cox05}). Cold atomic clouds transform to molecular clouds
if their volume density (to produce molecules fast enough) and
column density (to self-shield molecules against photodissociation)
are sufficiently high (e.g., \citealt{elm93,dra96}), as in, for
instance, shocks in turbulent flows \citep{glo07,glo09}. That the
star-forming molecular clouds strongly correlate with spiral arms
(e.g., \citealt{sta79,sol85,ken97,zim04,she07}) suggests that spiral
shocks too should trigger phase transitions from warm and diffuse to
cold and dense conditions.

Effects of TI on spiral shocks were first studied by \citet{shu73}, who
calculated one-dimensional shock profiles consisting of two stable
phases in equilibrium.  Although they allowed for
phase transitions, they assumed instantaneous thermal equilibrium,
which precluded the existence of transitory thermally-unstable
gas in their calculations.
Using direct time-dependent numerical simulations including ISM heating
and cooling,
\citet{tub80} and \citet{mar83} found that spiral shocks trigger
phase transitions if the initial density is large enough.
Because of strong numerical diffusion associated with insufficient
resolution, however, they were unable to capture TI in the postshock
transition zone, which is the thermally unstable regime.

In \citet[hereafter Paper II]{kko08}, we used high-resolution
one-dimensional simulations to study dynamical and thermodynamical
evolution of gas flows across spiral arms with ISM heating, cooling,
and thermal conduction. We found that even with TI, a quasi-steady
state develops with the following recurring cycle:  both warm and
cold phases in the interarm region are shocked and immediately
transform to denser cold gas in the arm, which subsequently evolves
to be TI-unstable due to postshock expansion in a transition zone,
and separates back into warm and cold phases. For a model with the
initial number density of $2\pcc$, the gas stays in the arm,
transition, and interarm zones for 14\%, 22\%, and 64\% of the
arm-to-arm crossing time, respectively. The gas mass in  the
TI-unstable temperature range was $\sim25-30\%$ of the total, and
mostly located in the transition zone, suggesting that postshock
expanding flows are important for producing intermediate-temperature
gas. Paper II also found that TI in association with one-dimensional
spiral shocks can drive random gas motions at $\sim 1.5\kms$ in the
interarm and transition zones; these velocities are $\sim5-7$ times
larger than those from pure TI alone (e.g., \citealt{kri02,pio04}).

In this paper, we extend the one-dimensional models of Paper II into
two dimensions, in order to study the effect of vertical disk
stratification on the dynamics and structure of multiphase galactic
spiral shocks. The current work also extends the
vertically-stratified isothermal models  considered in Paper I to
include the effects of the ISM heating and cooling. Our key
objective is to find how the flapping motions of spiral shocks
inherent in stratified disks interact with multi-phase gas produced
by TI, to change the shock structure and drive random gas motions in
each phase. We also study the internal properties of clouds that
form in self-gravitating models. Although \citet{dob07,dob_bon08}
and \citet{dob08b} ran SPH simulations to study shock structure and
cloud formation in three dimensions, they used pre-determined cold
and warm particles and did not allow the transitions between them.
\citet{dob08c} included ISM heating and cooling in the energy
equation and thus handled TI self-consistently, focusing on the
formation of molecular clouds in spiral shocks.  Using grid-based
three-dimensional simulations, \citet{wad08} studied dynamics of
galactic gas flows under the influence of self-gravity, a spiral-arm
potential, radiative cooling, star formation, and energy feedback
from supernova explosions. Although these three-dimensional global
models are of course more realistic, our local models are useful for
studying the detailed dynamics of spiral shocks at high spatial
resolution, and allow us to isolate each effect of the physical
processes involved.

This paper is organized as follows.
In \S2, we describe the basic equations we solve and specify the model
parameters.
In \S3, we present the results of one-dimensional solutions
for vertical disk equilibria including heating and cooling,
also providing simple analytic expressions for the mass fractions and
scale heights.
In \S4, we present the overall evolution, structure, and
statistical properties of spiral shock flows with TI in stratified
disks, based on the results of two-dimensional simulations.
The effect of self-gravity is discussed in \S5.
In \S6, we summarize our results and discuss their implications.

\section{Numerical Methods}

The local formulation used in the present study is similar to that
in Papers I and II.  In this section, we explain our numerical
method and model parameters.

\subsection{Basic Equations}\label{sec:eqn}

We consider a local region centered on a spiral arm that is assumed
to be tightly wound with a pitch angle $\sin i \ll 1$ and
rotating at a constant pattern speed $\Omega_p$.  We set up a local
Cartesian frame, centered at the position $(R, \phi,
z)=(R_0,\Omega_p t, 0)$, that corotates with the spiral arm.  The $x$--
and $y$--axes of the local frame are aligned in the midplane
parallel and perpendicular to the local arm segment, while $z$--axis
points the direction perpendicular to the galactic plane
(\citealt{rob69}; Paper I). Our simulation domain is a
two-dimensional rectangular region with size $L_x\times L_z$ in the
$x$--$z$ plane (hereafter XZ plane). We assume that all physical
variables are independent of $y$ (quasi-axisymmetric), while
allowing nonzero velocity in the $y$-direction in order to handle
epicycle motions associated with galactic rotation self-consistently.

In this local frame, the galactic differential rotation is translated
into the background velocity as
\begin{equation}\label{eq:vel0}
\vel_0=R_0(\Omega_0-\Omega_p)\sin i\xhat+
[R_0(\Omega_0-\Omega_p)-q_0\Omega_0 x]\yhat,
\end{equation}
where $\Omega_0=\Omega(R_0)$ and $q_0\equiv -(d\ln\Omega/d\ln R)|_{R_0}$
denotes the local shear rate in the absence of the spiral potential
\citep{kim02,kim06}.
Under the local approximation (i.e., $L_x \ll R_0$ and $|v| \ll R_0\Omega_0$),
the equations of hydrodynamics expanded in the local frame are
\begin{equation}\label{eq:cont}
 \frac{\partial\rho}{\partial t}+\nabla\cdot(\rho \vel_T)=0,
\end{equation}
\begin{equation}\label{eq:mom}
 \frac{\partial\vel_T}{\partial t}+\vel_T\cdot\nabla\vel_T=
 -\frac{1}{\rho}\nabla P -q_0\Omega_0 v_{0x}\yhat-2\mathbf{\Omega}_0\times
 \vel-\nabla(\Phi_s +\Phi_{\rm ext}),
\end{equation}
\begin{equation}\label{eq:energy}
 \frac{\partial e}{\partial t} +\vel_T\cdot\nabla e =
 -\frac{\gamma}{\gamma-1}P\nabla\cdot\vel_T -\rho\mathcal{L},
\end{equation}
\begin{equation}\label{eq:poisson}
\nabla^2\Phi_s=4\pi G\rho,
\end{equation}
\citep[e.g.,][]{rob69, shu73, kim06},
where
$\vel_T\equiv \vel_0+\vel$ is the total velocity in the local frame,
$\Phi_s$ is the self-gravitational potential of the gas,
$\Phi_{\rm ext}$ is the external stellar potential,
and $\rho\mathcal{L}(\rho, T)$ is the net cooling function.
Other symbols have their usual meanings.
We adopt an ideal gas law $P=(\gamma-1)e$ with $\gamma=5/3$.

The external stellar potential $\Phi_{\rm ext}$ varies in both the $x$-- and
$z$--directions and is separable into two parts as
\begin{equation}\label{eq:eph}
\Phi_{\rm ext}=2\pi G \rho_* z^2
+\Phi_{\rm sp}\cos\left(\frac{2\pi x}{L_x}\right),
\end{equation}
where $\rho_*$ is the unperturbed midplane stellar density, $\Phi_{\rm sp}$ is the
amplitude of the spiral potential,
and $L_x =2\pi R_0\sin i/m$ is the arm-to-arm separation
for an $m$-armed spiral. The first term in equation (\ref{eq:eph})
adopts a constant density $\rho_*$ for the stellar disk vertically; this is a
reasonable assumption given that most of the gas is located within
one stellar scale height of the midplane.  The second term is a local
analog of a logarithmic spiral potential considered in \citet{rob69}
and \citet{shu73}.  To parametrize the spiral arm strength, we employ
the dimensionless parameter
\begin{equation}\label{eq:F}
F\equiv \frac{m}{\sin i}\left(\frac{|\Phi_{\rm sp}|}{R_0^2\Omega_0^2}\right),
\end{equation}
corresponding to the ratio of the maximum force due to the spiral
potential to the the mean radial gravitational force
\citep{rob69}.

The interstellar gas is subject to the net cooling $\rho\mathcal{L}
\equiv n^2\Lambda[T]-n\Gamma$, where $n=\rho/(\mu m_{\rm H})$ is the
gas number density with $\mu=1.27$ being the mean molecular weight
per particle.  For the cooling rate of the diffuse ISM,
we take the fitting formula suggested by \citet{koy02},
\begin{eqnarray}\label{eq:cool}
 \Lambda(T)&=&2\times 10^{-19}\exp\left(\frac{-1.184\times10^5}{T+1000}\right)\nonumber\\
 &+&2.8\times10^{-28}\sqrt{T}\exp\left(\frac{-92}{T}\right)
{\rm \;erg} \cm^3 \;{\rm s^{-1}},
\end{eqnarray}
with temperature $T$ in degrees Kelvin.

For the gas heating function, we consider two different cases:
(1) a constant heating rate $\Gamma = \Gamma_0 = 2.0\times10^{-26}\ergs$
and (2) a density-dependent heating rate
\begin{equation}\label{eq:heat}
 \Gamma = \Gamma_0\exp(n/ n_0)^3,
\end{equation}
with $n_0=20\pcc$. The first, uniform heating rate corresponds to the
mean heating rate due to the
photoelectric effect on small grains and PAHs, photodissociations
of hydrogen molecules, and ionizations by cosmic rays and X-rays
(e.g., \citealt{koy02}). The second, density-modified heating rate that
increases stiffly with $n$ is to treat the effect of star
formation feedback in a very simple way.  Without such a
feedback, high-density clouds produced inside spiral arms in our
self-gravitating models would undergo catastrophic collapse,
preventing further computation.  In the real ISM,
gravitational collapse leads to new-born stars which will in turn
disperse the parent clouds by injecting radiative and mechanical energies.
Investigating the details of
star formation and consequent feedback processes is a very active
current research area.
Most previous work has adopted simplified feedback prescriptions that depend
on gas-consumption rate, star-formation efficiency, type of energy injection,
etc., with considerable uncertainties in the parameters
(e.g., \citealt{spr05,jou06,she08,ko09a}).
More realistic feedback prescriptions will be considered in
a subsequent paper.

\begin{figure}
 \epsscale{1.}
 \plotone{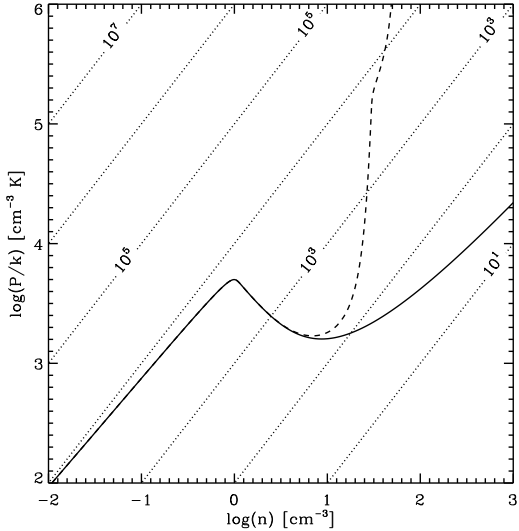}
 \caption{
 Thermal equilibrium curves in the density--pressure plane.
 Solid and dashed lines correspond to the uniform heating rate
 and the density-modified heating rate, respectively,
 while dotted lines indicate isotherms with $T$ in Kelvin.
 \label{fig:np0}
 }
\end{figure}

Figure~\ref{fig:np0} plots the equilibrium cooling curves in the
density \emph{vs}.\ pressure plane.  The solid line corresponds to
the uniform heating rate, while the dashed curve is for the modified
heating rate.  The dotted lines indicate isotherms.  The modified
heating rate changes the equilibrium curve dramatically only for
high-density gas, while making a negligible difference for
low-density material.  The equilibrium pressure has a local maximum
$P_{\rm max}/\kb=5.0 \times 10^3 \Kel\pcc$ at $n_{\rm crit,1}=1.0
\pcc$ for both heating rates, while attaining a local minimum at
$n_{\rm crit,2}=8.7 \pcc$ with $P_{\rm min}/\kb= 1.6 \times
10^3\Kel\pcc$ for the constant heating rate, and at $n_{\rm
crit,2}=6.9 \pcc$ with $P_{\rm min}/\kb= 1.7 \times 10^3\Kel\pcc$
for the modified heating rate. Under the constant heating rate,  the
gas temperature along the equilibrium curve is a monotonically
decreasing function of density, although it is insensitive to $n$ at
the low-density end with $n<n_{\rm crit,1}$. This is not the case
under the density-modified heating rate where the equilibrium
temperature increases with density when $n>n_{\rm
crit,3}=(1/3)^{1/3}n_0=13.9 \pcc$ in order to model feedback. We
thus classify the gas into two components based on its density
rather than temperature: \emph{rarefied} component if $n<n_{\rm
crit,1}$ and \emph{dense} component if $n>n_{\rm crit,1}$. Note that
thermally-unstable gas with $n_{\rm crit,1}<n< n_{\rm crit,2}$
belongs to the dense phase according to our classification.

As equation (\ref{eq:energy}) indicates, we explicitly ignore the
effect of thermal conduction in the present work.  Paper I found that
large translational motions in a finite difference scheme give rise
to numerical diffusion that tends to suppress the growth of TI,
similarly to thermal conduction.  The amount of numerical
conductivity in our models is typically $\mathcal{K}_n = 10^9 \ergs
\cm^{-1}\Kel^{-1}$ for the background velocity $v_{0x}=13\kms$, grid
spacing $\Delta x = 2.5\pc$, and the perturbation wavelength
$\lambda = 20\pc$.  Inclusion of physical conductivity larger
than $\mathcal{K}_n$ would resolve the wavelengths of the most
unstable TI.  But, this would in turn reduce the timestep greatly,
making computation essentially unpractical.\footnote{We have run some
simulations by including density-dependent thermal conductivity
$\mathcal{K}_n = 10^8\ergs\cm^{-1}\Kel^{-1}
(1+0.05\pcc/n)^{-1}$ \citep{ko09a}, and confirmed that this level of
thermal conduction does not make a significant difference in the
results.}
We note that by neglecting the thermal conduction term
in the energy equation, some of our results may depend on the
numerical resolution, although the mass fractions appear to be
insensitive to the resolution (Paper I).

\subsection{Model Parameters \& Numerical Methods \label{sec:model}}

\begin{deluxetable*}{lccccccc}
\tabletypesize{\footnotesize} \tablecaption{Models Without Spiral
Arms ($F=0\%$) \label{tbl:noarm}} \tablewidth{0pt} \tablehead{
\colhead{Model\tablenotemark{a}} &
\colhead{$\Sigma_0\; ({\rm M_\odot\;pc^{-2}})$} &
\colhead{$f_D \;(\%)$} &
\colhead{$f_R \;(\%)$} &
\colhead{$H_{ D}\;(\rm pc)$} &
\colhead{$H_{ R}\; (\rm pc)$} &
\colhead{$H_{\rm ave}\; (\rm pc)$}  \\
\colhead{(1)} &
\colhead{(2)} &
\colhead{(3)} &
\colhead{(4)} &
\colhead{(5)} &
\colhead{(6)} &
\colhead{(7)}
}
\startdata
  NU.S02 &      2 &      0 &    100 &      0 &    126 &    126  \\
  NU.S05 &      5 &      0 &    100 &      0 &    119 &    119  \\
  NU.S10 &     10 &     71 &     29 &      2 &    125 &     67  \\
  NU.S20 &     20 &     86 &     14 &      4 &    127 &     48  \\
  NM.S02 &      2 &      0 &    100 &      0 &    126 &    126  \\
  NM.S05 &      5 &      0 &    100 &      0 &    119 &    119  \\
  NM.S10 &     10 &     69 &     31 &      4 &    125 &     70  \\
  NM.S20 &     20 &     85 &     15 &      9 &    130 &     50  \\
  SM.S02 &      2 &      0 &    100 &      0 &    121 &    121  \\
  SM.S05 &      5 &      0 &    100 &      0 &    107 &    107  \\
  SM.S10 &     10 &     82 &     18 &      4 &    100 &     43  \\
  SM.S20 &     20 &     94 &      6 &      7 &     84 &     21
\enddata

\tablenotetext{a}{The prefixes NU refers to the non-self-gravitating
models with the uniform heating rate, NM for the
non-self-gravitating models with the modified heating rate, and SM
for the self-gravitating models with the modified heating rate.}

\end{deluxetable*}

\begin{deluxetable*}{lcccccccc}
\tabletypesize{\footnotesize} \tablecaption{Models With Spiral Arms
($F=5\%$) \label{tbl:arm}} \tablewidth{0pt} \tablehead{
\colhead{Model} &
\colhead{$\Sigma_0\; ({\rm M _\odot\;pc^{-2}})$} &
\colhead{$f_D\;(\%)$} &
\colhead{$f_R\;(\%)$} &
\colhead{$H_{ D}\; (\rm pc)$} &
\colhead{$H_{ R}\; (\rm pc)$} &
\colhead{$H_{\rm ave}\; (\rm pc)$} \\
\colhead{(1)} &
\colhead{(2)} &
\colhead{(3)} &
\colhead{(4)} &
\colhead{(5)} &
\colhead{(6)} &
\colhead{(7)}
}
\startdata
  NU.S02 &      2 &     11 &     89 &     25 &    129 &    122  \\
  NU.S05 &      5 &     62 &     38 &     10 &    130 &     81  \\
  NU.S10 &     10 &     81 &     19 &      7 &    130 &     57  \\
  NM.S02 &      2 &     12 &     88 &     28 &    130 &    122  \\
  NM.S05 &      5 &     60 &     40 &     20 &    132 &     84  \\
  NM.S10 &     10 &     81 &     19 &     25 &    135 &     64  \\
  SM.S02 &      2 &     26 &     74 &     14 &    124 &    107  \\
  SM.S05 &      5 &     91 &      9 &     21 &    121 &     42  \\
  SM.S10 &     10 &     95 &      5 &     43 &    123 &     51
\enddata

\tablecomments{Model name prefixes are as in Table \ref{tbl:noarm}.}

\end{deluxetable*}

We consider a simulation box located near the Solar neighborhood at
galactocentric radius of $R_0=8\kpc$.  We adopt the galactic
rotational velocity of $R_0\Omega_0 = 208 \kms$ with a flat rotation
curve ($q_0=1$). The corresponding angular velocity is
$\Omega_0=26\kms\kpc^{-1}$, and orbital period is
$\torb\equiv2\pi/\Omega_0=2.4\times10^8\yr$, which we use as the
time unit in our presentation. For spiral arm parameters, we take
pattern speed $\Omega_p=0.5\Omega_0$, pitch angle $\sin i = 0.1$,
and azimuthal wavenumber $m=2$. The corresponding arm-to-arm
separation is $L_x=2\pi R_0 \sin i/m=2.5\kpc$, which is set equal to
the size of the simulation box along the $x$--direction.  We fix the
spiral arm strength to $F_0= 5\%$.

Our initial gaseous disks, in the absence of the spiral-arm perturbations,
are taken to be isothermal and in vertical hydrostatic equilibrium under
the linear stellar gravity $g_z =-4\pi G\rho_*z$ (see eq.\ [\ref{eq:eph}]).
The corresponding density distribution is a Gaussian  profile
\begin{equation}\label{eq:Gauss}
\rho(z)=\rho_0 \exp\left(-\frac{z^2}{2h_g^2}\right),
\end{equation}
with a scale height
\begin{eqnarray}\label{eq:scalehR}
h_g&=&\frac{c_R}{(4\pi G\rho_*)^{1/2}}\nonumber\\
&=& 128\pc\left(\frac{c_R}{7\kms}\right)\left(\frac{\rho_*}
{0.056\Msun\pc^{-3}}\right)^{-1/2},
\end{eqnarray}
where $c_R=7\kms$ is the isothermal sound speed of the initial disks
and $\rho_* = 0.056\Msun\pc^{-3}$  is the stellar density near the
solar neighborhood \citep{hol00}. We take $L_z=7.5h_g = 960\pc$ as the
vertical size of the simulation domain (i.e., $|z|\leq L_z/2$).

Tables~\ref{tbl:noarm} and \ref{tbl:arm} summarize the model parameters
and some simulation outcomes for models with and without spiral
potential perturbations, respectively.
Column (1) labels each run. The prefixes NU and NM stand for
non-self-gravitating models (``N'') with the uniform heating rate
(``U'') and the
modified heating rate (``M''), respectively, while the prefix SM indicates
self-gravitating models (``S'') with the modified heating rate (``M'').
As will be discussed below, column (2) gives the initial gas surface density $\Sigma_0$.
Columns (3) and (4) give the mass fractions,
$f_a \equiv \langle \int \rho_a dxdz /\int \rho dxdz\rangle $
(with $a=D$ or $R$), of dense and rarefied components, respectively.
Here, the angle brackets $\langle~\rangle$ denote a time average
over $t/\torb=5-8$ for non-self-gravitating models and
over $t/\torb=8-11$ for self-gravitating models.
Columns (5) and (6) give the scale heights,
$H_a \equiv \langle \int \rho_a z^2 dxdz /
\int \rho_a dxdz \rangle^{1/2}$, of the dense and rarefied
components, respectively.
Column (7) gives the average scale height of the whole gas
$H_{\rm ave} \equiv \left(f_D H_D^2 + f_R H_R^2\right)^{1/2}$.

We integrate the time-dependent partial differential equations
(\ref{eq:cont})--(\ref{eq:poisson}) using a modified version of the
Athena code \citep{gar05,gar08,sto08,sto09}. Athena employs a
single-step, directionally unsplit Godunov scheme for
magnetohydrodynamics in multispatial dimensions. Among the various
schemes contained in Athena, we take a piecewise linear method for
spatial reconstruction, HLLE Riemann solver to compute the fluxes
\citep{har83,ein91}, and van Leer algorithm for directionally
unsplit integration \citep{sto09}. Since our simulations involve
strong shocks for the dense medium (with typical Mach numbers
$\sim 7-10$), we adopt the first order flux correction when the net
mass flux out of a cell exceeds the initial mass of the cell in order to
avoid an occurrence of negative density (see, e.g., \citealt{lem09}).
Our models employ a $1024\times 512$ zones over
the simulation box, corresponding to the
resolution of $\Delta x= 2.4\pc$ and $\Delta z= 1.9\pc$.

We adopt the shearing-periodic boundary condition at the
$x$--boundaries \citep{haw95}.  In the $z$--direction, we use the
outflow condition for the velocity and the vacuum condition for the
gravitational potential (e.g., \citealt{ko09a}). For the density and
pressure at the $z$--boundaries, we linearly extrapolate the
logarithmic density, while keeping temperature fixed, whenever
$d\rho/dz<0$.  This produces a balance between the vertical pressure
gradient and the gravitational source term at the boundaries,
similarly to the ``conducting'' boundary in \citet{par05}.  When
$d\rho/dz>0$, on the other hand, we switch to the continuous
boundary condition for the density and pressure to reduce artificial
mass inflow due to the extrapolation. Under our boundary conditions,
the gas can freely escape from the vertical boundary; we have
checked that the total mass is nonetheless conserved within $2\%$
for all models. Because of the very short cooling time, energy
updates from the net cooling functions are made implicitly based on
Newton-Raphson iterations \citep{pio04}. To solve for the
gravitational potential in our simulation domain, we adopt a method
introduced by \citet{ko09a} which, by using the fast Fourier
transform technique, is much more efficient than a hybrid method
involving Green's functions (e.g., \citealt{kos02})

\section{Vertical Equilibria Without Spiral Arms}\label{sec:vequil}

While our main objective is to investigate the overall dynamics and
structure of spiral shocks in vertically stratified disks under the
influence of TI, in this section we focus on the quasi-static
vertical equilibria with heating and cooling in the absence of the
spiral arm potential (i.e., $F=0$). This allows us to study the
effect of TI on vertical disk structure.  We run one-dimensional
simulations with physical quantities varying only with $z$.  We
consider an initially-isothermal disk with $\Sigma_0=2,$ 5, 10, or
$20\Surf$, and evolve it subject to TI.

\begin{figure}
 \epsscale{1.}
 \plotone{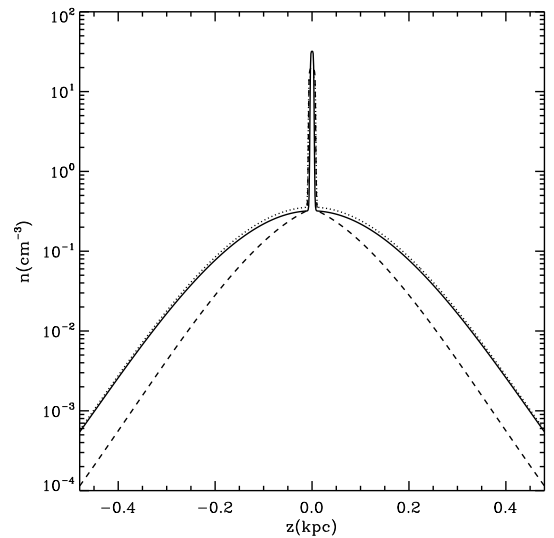}
 \caption{Distributions of number density along the vertical direction
 for one-dimensional non-self-gravitating
 models NU.S10 (\emph{solid}), NM.S10 (\emph{dotted}),
 and self-gravitating model
 SM.S10 (\emph{dashed}) without spiral-arm potential perturbations.
 The midplane densities are $n(0)=32$, 15, and $20\pcc$ for models NU.S10,
 NM.S10, and SM.S10, respectively.
 The interface between the midplane dense layer and the surrounding
 rarefied medium occurs at almost the same density $n_{\rm trans}\sim0.25-0.35\pcc$,
 corresponding to $P_{\rm trans}/\kb\sim2000-2200\pcc\Kel$.
 \label{fig:zprof0}
 }
\end{figure}

For disks with large surface density (models with $\Sigma_0 \geq
10\Surf$), TI grows rapidly ($\ll \torb$), transforming the
initially constant-temperature gas into thermally bistable phases.
The cold, dense gas falls toward the midplane to form a dense slab,
while the warm, rarefied gas rises up buoyantly.  The infall is
supersonic relative to the dense medium.  At early time, the dense
slab surrounded by the upper rarefied gas undergoes vertical
expansions and contractions a few times.  As the kinetic energy
dissipates through shocks at the interfaces, the whole configuration
evolves toward vertical hydrostatic equilibrium typically within
$\sim 0.6\torb$. Figure~\ref{fig:zprof0} shows density profiles for
S10 models with $\Sigma_0=10\Surf$. Solid and dotted lines are for
non-self-gravitating NU and NM models, respectively, while the
dashed line is for the self-gravitating SM models. The difference
between models NU.S10 and NM.S10 is not significant since the
maximum midplane density is not much larger than $n_0 = 20\pcc$,
below which the heating rate is almost density-independent. For
model SM.S10, self-gravity compresses the midplane slab further at
the expense of the rarefied medium at $|z|>H_D$. Nevertheless, the
phase transition between dense and rarefied components turns out to
occur at almost the same density $n_{\rm trans} \sim 0.25-0.35\pcc$,
corresponding to the transition pressure $P_{\rm trans}/\kb \sim
2000-2200\pcc\Kel$ for all models that are unstable to TI.
Note that $P_{\rm trans}/\kb$ is above the minimum pressure for a cold
medium with our adopted cooling and heating functions,
$P_{\rm min}/\kb =1600-1700 \pcc\Kel$.

\begin{figure}
 \epsscale{1.}
 \plotone{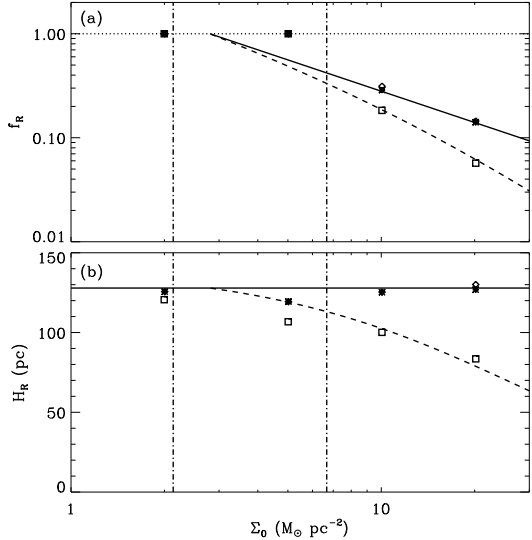}
 \caption{(\emph{a}) Mass fraction and (\emph{b}) scale height
 of the rarefied component as functions of total surface density
 $\Sigma_0$ from one-dimensional simulations without spiral arms.
 Symbols represent the numerical results for non-self-gravitating
 models NU (\emph{asterisks}),
 NM (\emph{diamonds}), and self-gravitating models SM (\emph{squares}).
 Solid and dashed curves are the theoretical estimates for two-phase
 equilibrium with and without self-gravity, respectively, for which
 $P_R(0)/k=P_{\rm trans}/\kb=2100 \Kel \pcc$ and $c_R=7\kms$ are adopted.
 Vertical dot-dashed lines mark $\Sigma_{\rm min}=2.1\Surf$
 and $\Sigma_{\rm max}=6.7\Surf$; for
 $\Sigma_{\rm min}<\Sigma < \Sigma_{\rm max}$,
 both single-phase and two-phase equilibria are possible.
 \label{fig:mf0}
 }
\end{figure}

Once vertical hydrostatic equilibrium is attained, we measure the
mass fractions $f_D$ and $f_R$, and the scale heights $H_D$ and
$H_R$ of the dense and rarefied components, respectively; these
values are listed in Table~\ref{tbl:noarm}. Figure~\ref{fig:mf0}
plots $f_R$ and $H_R$ as functions of the initial surface density
$\Sigma_0$.  The results of NU and NM models are denoted by
asterisks and diamonds, respectively, while open squares are for SM
models. Models with low surface density ($\Sigma_0=2\Surf$) do not
experience TI and thus establish a single-phase equilibrium
consisting only of the rarefied medium. Since the warm gas is nearly
isothermal at $c_R\approx7\kms$ and self-gravity is weak in these
models, the equilibrium density profiles in low $\Sigma$ models are
approximately given by equation (\ref{eq:Gauss}), with surface
density $\Sigma= \rho_0 h_g \sqrt{2 \pi}=\rho_0 c_R /\sqrt{2 G
\rho_*}$.  Since the midplane pressure $P_R(0)=c_R^2\rho_0$ of the
rarefied gas cannot exceed $P_{\rm max}$ along the thermal
equilibrium curve shown in Figure~\ref{fig:np0}, the surface density
for a single-phase equilibrium with only a rarefied component is
constrained to be less than $\Sigma_{\rm max}=P_{\rm
max}/(2G\rho_*c_R^2)^{1/2} =6.7\Surf$.
Similarly, the condition $P_R(0)=P_{\rm min}$ yields
$\Sigma_{\rm min}=2.1 \Surf$ as the minimum surface density
for a two-phase equilibrium in which dense and rarefied components
coexist. The two vertical dot-dashed lines in Figure~\ref{fig:mf0}
mark $\Sigma_{\rm min}$ and $\Sigma_{\rm max}$. Below $\Sigma_{\rm
min}$, only a rarefied phase is possible, whereas above $\Sigma_{\rm
max}$, both dense and rarefied phases must be present.

For $\Sigma_{\rm min}<\Sigma_0<\Sigma_{\rm max}$, both single (rarefied)
phase and two-phase equilibria can be realized.  Which type of
equilibrium emerges depends of course on the initial disk conditions.
In the case of our models with $\Sigma_0=5\Surf$,
the initial midplane density and pressure are $n(0)= 0.5\pcc$
and $P(0)/\kb=3770\Kel\pcc$, smaller than than $n_{\rm crit,1}$ and
$P_{\rm max}/\kb$.
Since cooling and heating occur almost isobarically,
even the densest gas in these models is unable to overcome
$P_{\rm max}$ to turn into the dense component, for this case.

Figure~\ref{fig:mf0} also shows that for the models that reach a
two-phase equilibrium, self-gravity reduces the rarefied-gas fraction in mass
as well as its scale height compared to those in
non-self-gravitating counterparts.  Self-gravity also makes the density profile of
the rarefied component deviate significantly from a Gaussian profile.
A thin midplane dense slab, containing the majority of the gas mass,
exerts a uniform gravity on  the rarefied gas lying above it, providing an
additional confining force.  In the Appendix, we describe a simple way to
estimate $f_R$ and $H_R$ as functions of the total gas density,
assuming that the rarefied component can be characterized by
a fixed sound speed $c_R$ and that its self-gravity is negligible.
The resulting theoretical predictions, with $c_R=7\kms$,
for self-gravitating and non-self-gravitating cases are plotted in
Figure~\ref{fig:mf0} as dashed and solid curves, respectively,
These are overall in good agreement with the numerical results.
Small discrepancies between the theoretical and numerical values of
$H_R$ for disks with $\Sigma_0=5\Surf$ arise from the fact that
the rarefied gas in these models has larger midplane pressure
than in any other models.\footnote{
For example, the midplane pressure of the rarefied component is
$P_R(0)/\kb= 3500$ and $2500 \pcc\Kel$ for models
NM.S05 and NM.S10, respectively.}
In view of the thermal equilibrium curve shown in Figure~\ref{fig:np0},
this implies that the rarefied medium in S05 models is coldest,
corresponding to $c_R \simeq 6.3\kms$, making the scale height
smaller than the theoretical estimate based on $c_R=7\kms$.

\section{Non-self-gravitating Models}

Now we turn to our main theme, nonlinear gas flows with TI across
spiral arms in a stratified disk. In this section, we study
overall evolution, structure, and statistical properties such as density
and random velocity distributions of spiral shocks for non-self-gravitating
models. Effects of self-gravity will be discussed in the next section.

\subsection{Overall Evolution}

\begin{figure*}
 \epsscale{1.}
 \plotone{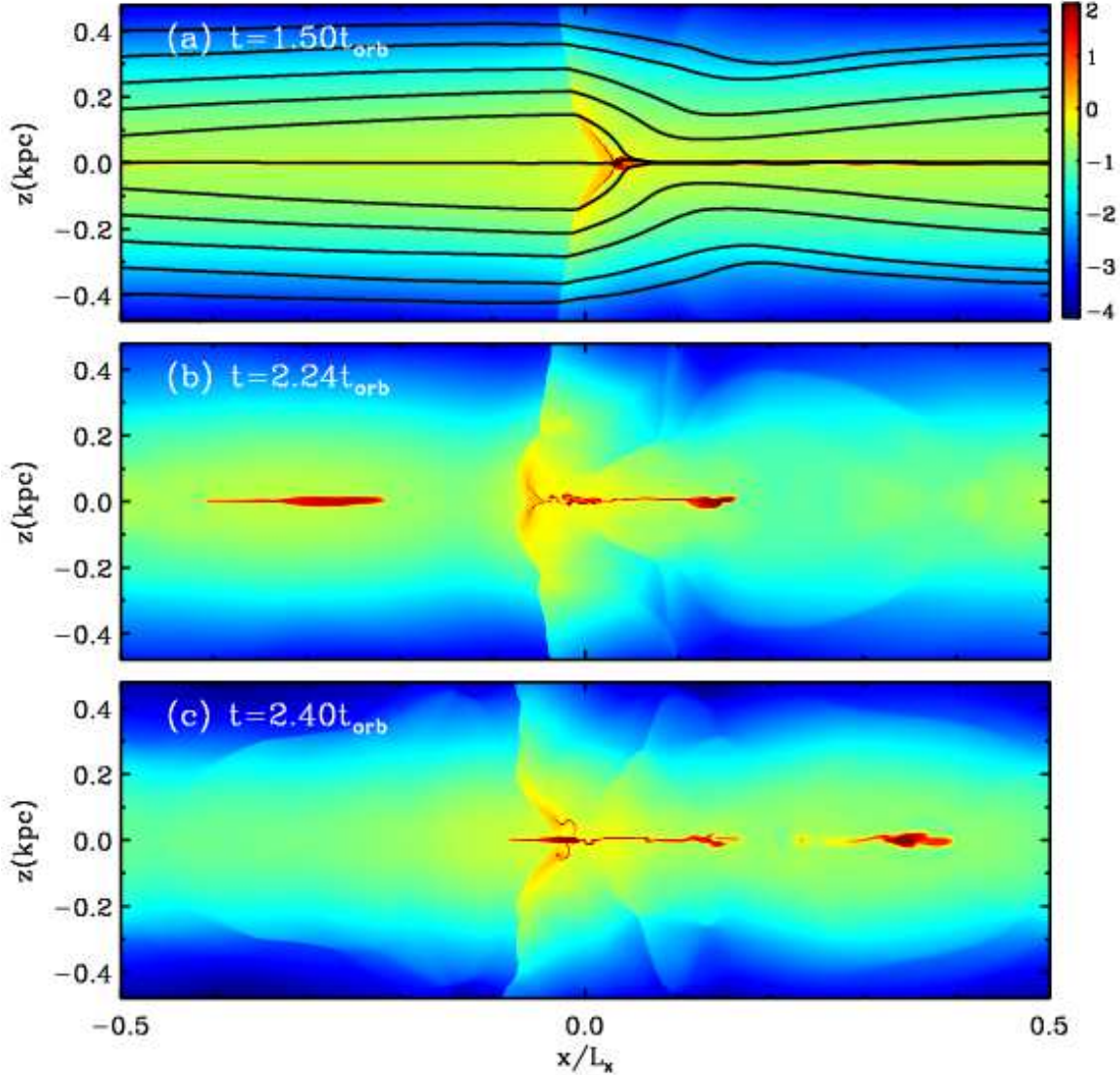}
 \caption{Density snapshots for model NU.S10 at
  $t/\torb=1.50$, 2.24, and 2.40.
  A few instantaneous streamlines are drawn as solid lines in (\emph{a}).
  The shock
  front alternates between convex (\emph{b}) and concave (\emph{c}) shapes,
  seen from the upstream direction, due to quasi-periodic flapping.
  Three dense condensations located near $x/L_x=-0.3$, 0, and 0.14 in
  (\emph{b}) have moved to $x/L_x=-0.12$, 0.14, and 0.35 in (\emph{c}),
  respectively.
  Colorbar labels $\log (n/1\pcc)$.
 \label{fig:na10}
 }
\end{figure*}

\begin{figure*}
 \epsscale{1.}
 \plotone{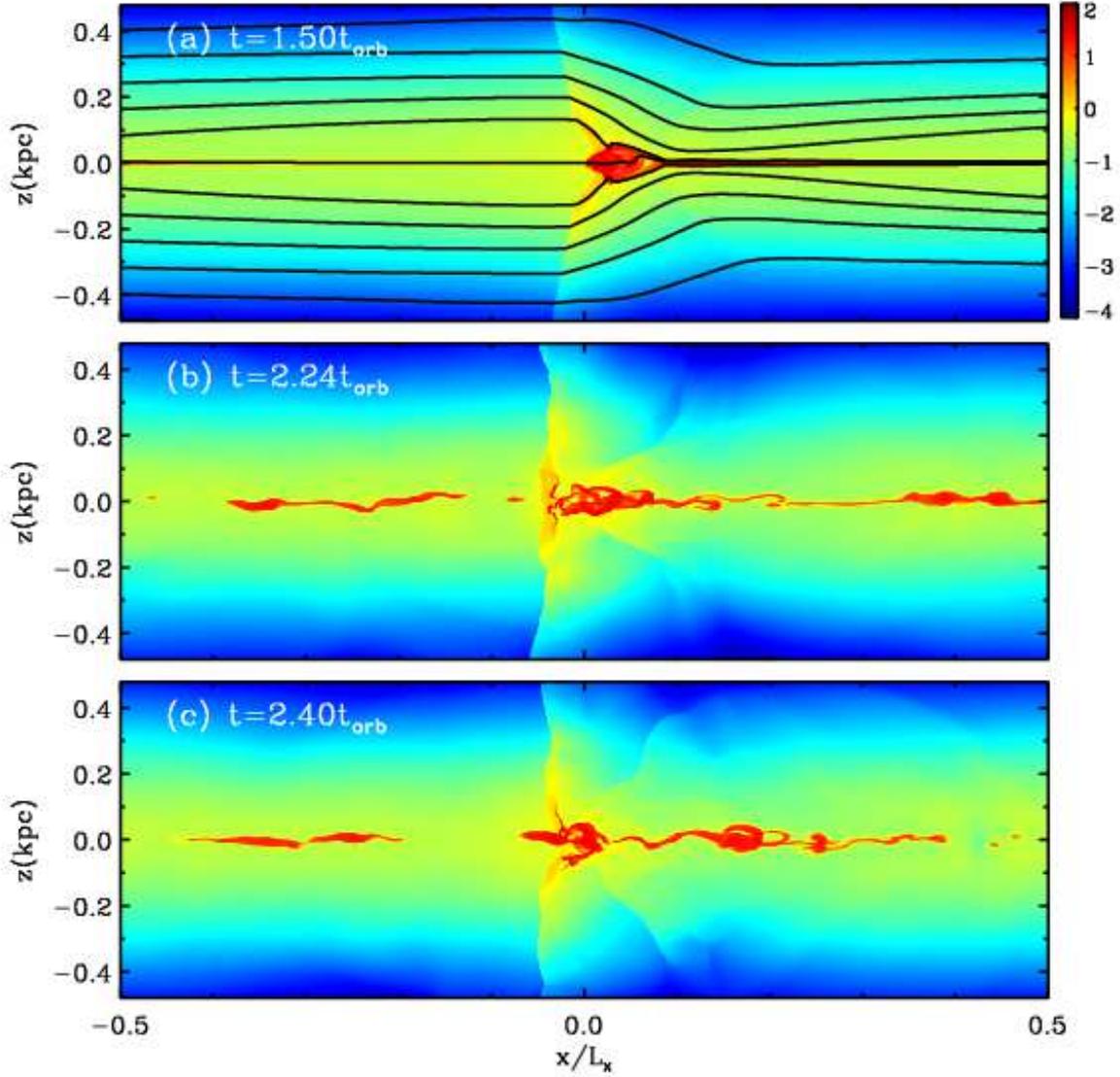}
 \caption{Density snapshots for model NM.S10 at
  $t/\torb=1.50$, 2.2, and 2.40.
  A few instantaneous streamlines are drawn as solid lines in (\emph{a}).
  Compared to model NU.S10, the density-modified heating rate thickens
  the dense midplane layer and prohibits the formation of dense
  condensations.
  Colorbar labels $\log (n/1\pcc)$.
 \label{fig:nb10}
 }
\end{figure*}

\begin{figure*}
 \epsscale{1.}
 \plotone{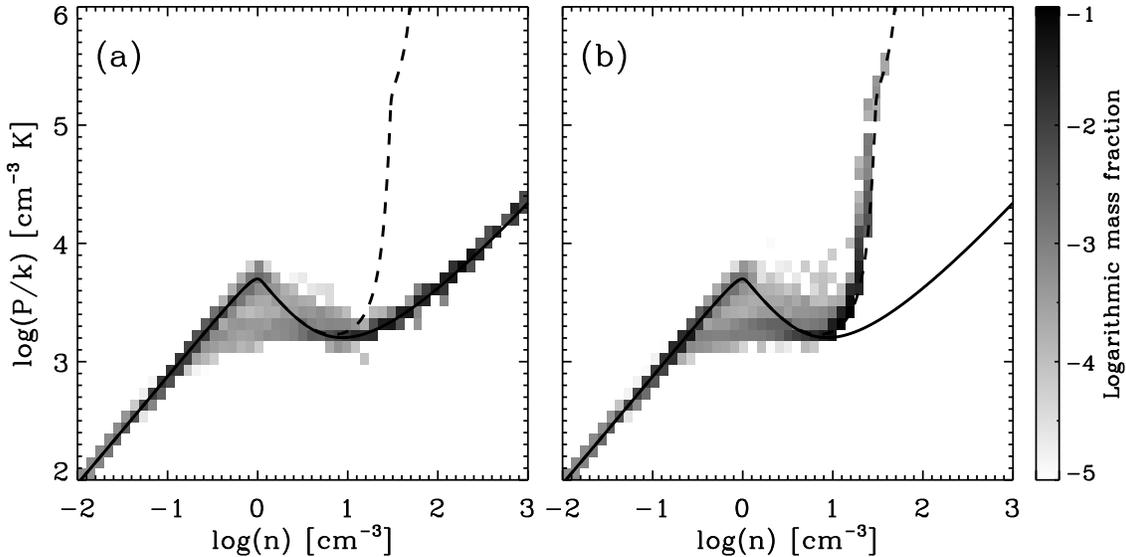}
 \caption{
 Distribution of gas in the density-pressure plane for
 models (\emph{a}) NU.S10 and (\emph{b}) NM.S10 at $t/\torb=2.4$,
 with grayscale indicating the mass fraction in logarithmic scale.
 The thermal equilibrium curves are the same as in Figure~\ref{fig:np0}.
 \label{fig:np}
 }
\end{figure*}

We begin by describing evolution of our fiducial models NU.S10 and
NM.S10 with $\Sigma_0=10\Msun\pc^{-2}$ that employ the uniform and
density-modified heating rates, respectively.  We slowly turn on the
spiral potential amplitude such that it attains full strength
$F_0=5\%$ at $t/\torb =1.5$. Snapshots of volume density in
logarithmic color scale at early epochs $t/\torb=1.50$, $2.24$, and
$2.40$ are shown in Figures~\ref{fig:na10} and ~\ref{fig:nb10},
respectively. Figure~\ref{fig:np} plots the gas distribution in the
$n$-$P/\kb$ plane for both models at $t/\torb=2.40$. Initially, the
disk is in hydrostatic equilibrium with a constant sound speed of
$c_R=7\kms$. Since the initial disk is out of thermal equilibrium,
it quickly evolves into a two-phase equilibrium configuration, as
explained in \S\ref{sec:vequil}.  As $F$ increases, both the dense
gas  near the midplane and the rarefied gas at high $|z|$ respond to
the growing spiral potential and form a shock front near the
potential minimum.

Since the gas flows at this time are fairly horizontal without much
vertical mixing, as evidenced by the  instantaneous streamlines
shown in Figures~\ref{fig:na10}\emph{a} and \ref{fig:nb10}\emph{a},
the shock profile at each height is very similar to those of
the one-dimensional cases studied in Paper II.
The shock strength and gas phase in the postshock regions depend on
the mean density and temperature at that height.  Near the midplane
at $|z|< H_D$ ($=7$ and $25\pc$ for models NU.S10 and NM.S10, respectively),
the dense slab is so cold that the shock is very strong with a typical
Mach number $\mathcal{M}\sim 10$, resulting in a far denser postshock
region.  In the high-$|z|$ regions ($|z|>130\pc$),
on the other hand, the gas is warm and has a low mean density ($<0.1\pcc$)
enough to remain warm even after the
shock compression. It is the mid-altitude rarefied medium
(at $H_D < |z| < 130\pc$) that is able to achieve a postshock pressure
larger than $P_{\rm max}$ and thus undergoes a phase transition to
the dense component after experiencing isobaric cooling
(\citealt{muf74,ino08}; Paper II).
Since the shock is stronger at lower $|z|$ in a stratified disk
and since a stronger shock tends to move downstream \citep[e.g.,][]{kim02},
the shock front when it first develops is naturally curved in the XZ plane.
Figures~\ref{fig:na10}\emph{a} and \ref{fig:nb10}\emph{a} show that
the shock front is concave when seen from the upstream direction,
with mean slopes of $|d x_{\rm sp}(z)/dz|\approx 0.83$ at $|z|<130 \pc$
and $0.13$ at $|z|>130 \pc$, where $x_{\rm sp}(z)$ is the shock
location at $z$.

The dense gas produced at the shock at moderate $z$ begins falling
toward the midplane under the influence of the external gravity as
it moves downstream. The reduction of the velocity in the direction
normal to the concave shock front also helps the downward motion of
the gas. On the other hand, the dense gas near the midplane has a
large pressure and thus slightly expands vertically after the shock.
The vertical expansion is more extreme in NM models than in NU
models. The falling gas collides with the expanding gas, reducing
the rising motion of the latter. The streamlines shown in
Figures~\ref{fig:na10}\emph{a} and \ref{fig:nb10}\emph{a} illustrate
these motions at early time.

The rarefied gas which crosses the shock at high $|z|$ also falls toward the
midplane as it follows galaxy rotation.  This builds up thermal
pressure at low $|z|$,  so the flow rebounds to high-altitude regions.
Since the period of vertical oscillation, $\sim
(G\rho_*)^{-1/2}$, is in general not commensurate with the interval
between arm crossings, the streamlines of the rarefied gas are not closed.
This causes the shock front to sway back and forth around its mean position
in the direction perpendicular to the arm (e.g., \citealt{kim06}; Paper I).
During the course of the flapping motions,
the shock front has a convex shape (seen from upstream)
when the postshock regions are maximally
compressed (Fig.\ \ref{fig:na10}\emph{b}), while it becomes
concave when the gas in the postshock regions is in full
vertical expansion (Fig.\ \ref{fig:na10}\emph{c}).
These flapping motions of the shock front,
alternating between convex and concave shapes,
occur quasi-periodically with
a period of $\sim 0.5\torb$ and have an amplitude of
$\Delta x/H_R \sim 1$ at $|z|=H_R$ ($=130\pc$ in model NU.S10).
The shock flapping motions are able to tap some of the kinetic energy
in galaxy rotation to supply random kinetic energy for the gas.
We will quantify the amplitudes of random gas motions driven by
flapping in \S\ref{sec:turb}.

One of the special features of galactic spiral shocks is that gas
experiences acceleration after the maximum shock compression,
forming a postshock expansion zone (e.g., \citealt{bal88,kim06};
Papers I \& II). Any parcel of gas becomes gradually less dense as
it moves downstream in the expansion zone. In model NU.S10, the
shock compression and subsequent cooling is so strong that the
shocked dense gas in  the midplane can reach $n>10^3\pcc$ (see also
model SC20 in Paper II). With such a large postshock density, this
gas can still remain dense, with $n>n_{\rm crit, 2}$, even after
emerging from the expansion zone located at $x/L_x\sim0-0.3$. This
TI-stable dense gas travels almost ballistically in the interarm
region, reenters the arm, and combines with other dense gas to
produce a few condensations. Figures~\ref{fig:na10}\emph{b} at time
$t/\torb=2.24$ shows three large condensations located at $x/L_x\sim
-0.3$, $0$, and $0.14$, which are stretched horizontally due to the
expanding background velocity.  The condensations move nearly
horizontally to appear at $x/L_x=-0.12$, 0.14, and 0.35 when
$t/\torb=2.40$ (Fig.~\ref{fig:na10}\emph{c}). In model NU.S10, the
dense condensations stay in the arm ($-0.05 < x/L_x < 0.05$)
for about $\sim 0.15\torb$, in the expansion zone
($0.05<x/L_x<0.25$) for about $\sim 0.30\torb$, and in the interarm
region for the remainder ($\sim 0.55\torb$) of the cycle. The width
and residence time of each zone are insensitive to the model
parameters.

With the density-modified heating rate, on the other hand, the postshock
dense gas in model NM.S10 has a moderate density ($\sim30-40\pcc$), so that
the postshock expansion is able to take it to the thermally unstable
regime ($n_{\rm crit, 1}<n<n_{\rm crit, 2}$).
Subsequently, the expanding dense gas suffers from TI
and separates back into dense and rarefied gas in the interarm region.
The large thermal pressure also prevents the formation of dense condensations
in this model.

\begin{figure}
 \epsscale{1.}
 \plotone{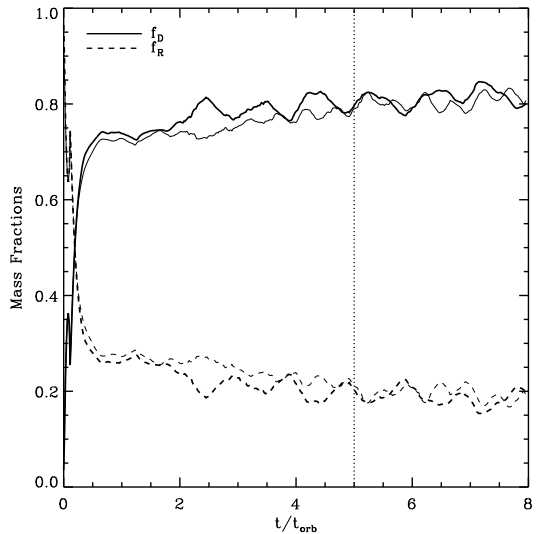}
 \caption{
 Mass fractions of the dense ($f_D$) and rarefied ($f_R$) components
 as functions of time for models NU.S10 ({\it thick}) and NM.S10 ({\it thin}).
 Initially, $f_D$ increases rapidly as the gas cools due to TI
 and flattens to $f_D\sim  0.7$ at $t/\torb\sim 0.6$ when hydrostatic
 equilibrium is attained before the effect of the spiral potential becomes
 significant.  The presence of
 the spiral arm at full strength increases this to a saturate value
 of $f_D\sim0.8$ at $t/\torb \simgt 5$.   The mass
 fractions of model NU (unmodified heating) and NM (modified heating)
 are quite similar.
 \label{fig:mf_tevol}
 }
\end{figure}

Figure~\ref{fig:mf_tevol} plots the temporal evolution of the mass
fractions of dense phase (\emph{solid lines}) and rarefied phase
(\emph{dashed lines}), respectively,  for models NU.S10 (\emph{thick
lines}) and NM.S10 (\emph{thin lines}). At early time, $f_D$
increases rapidly as the gas cools and collapses toward the midplane
to form a dense layer that bounces appreciably at $t/\torb\sim0.1$.
The mass fractions flatten at $t/\torb\sim 0.6$ when vertical
hydrostatic equilibrium is established, well before the effect of
the spiral potential becomes substantial. As the spiral potential
attains its full strength at $t/\torb=1.5$, $f_D$ increases slightly
due to the phase transition of the rarefied to dense phases
occurring at the shock fronts. Although the flows are fully
nonlinear with strong unsteady motions and phase transitions, there
is no noticeable secular variation in the mass fractions, which
remains at $f_D\sim 0.8$  after  $t/\torb=5$; the associated
temporal fluctuation amplitudes are about $6-9\%$
relative to the mean values. We thus conclude that in a statistical
sense, the spiral shocks in our models have reached a quasi-steady
state at $t/\torb>5$. Compared with models without spiral arms
discussed in \S\ref{sec:vequil}, the shock compression and
associated phase transitions decrease the rarefied gas fraction by
$46\%$ for S10 models.  In fact, all of the non-self-gravitating models
with spiral arms have comparable total surface density of rarefied gas,
$\sim 1.9 \Msun\pc^{-2}$,  lower than the value
$\Sigma_R= P_{\rm trans}/(c_R \sqrt{2 G \rho_*})\approx 2.8 \Msun\pc^{-2}$
that would be predicted using a uniform surface density.
Note that both NU and NM models have almost
the same dense and rarefied mass fractions since the modified
heating rate does not affect the rarefied medium much.

\begin{figure*}
 \epsscale{1.}
 \plotone{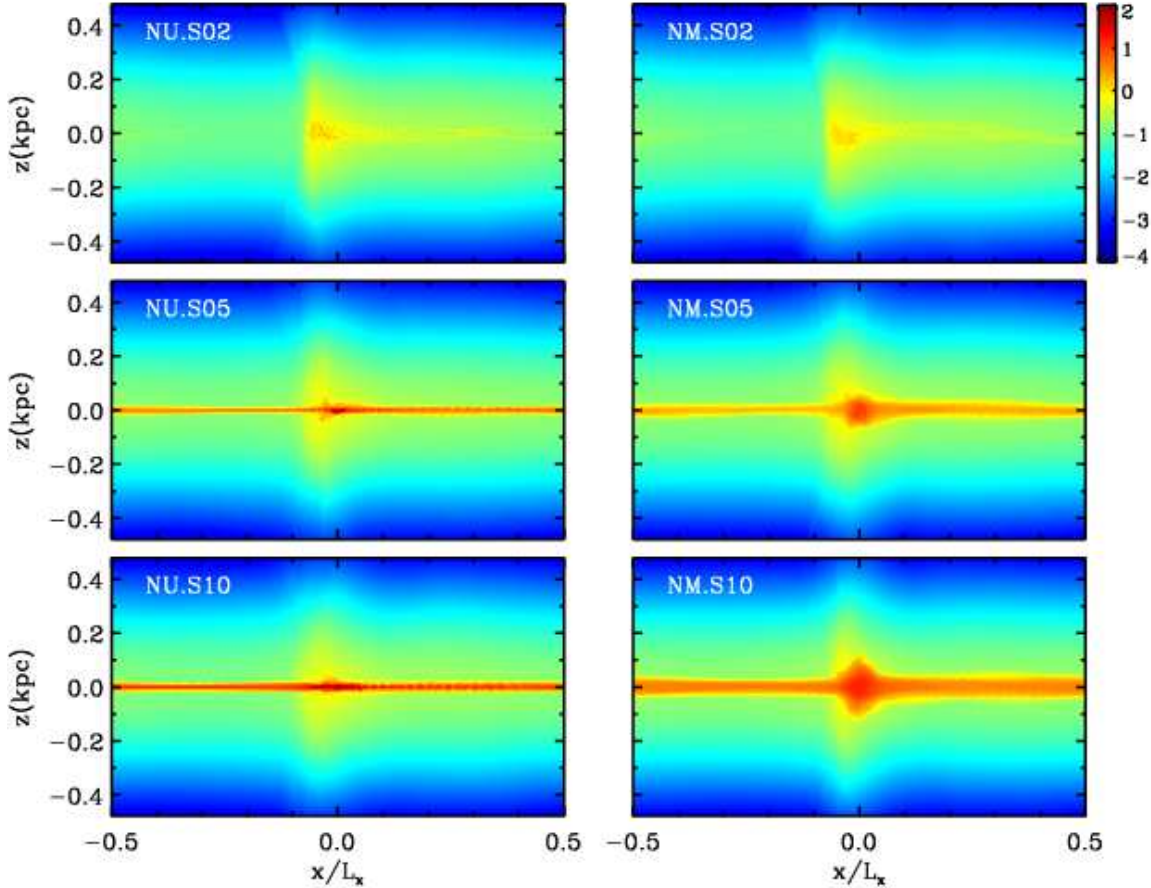}
 \caption{
 Density distributions
 averaged over $t/\torb=5$--8 of our non-self-gravitating models.
 S05 and S10 models contain midplane dense gas in both arm and interarm
 regions, while S02 models have the dense phase only in the arm regions.
 Compared to NU models, the arm regions in NM models are broader and
 thicker.
 Colorbar labels $\log (n/1\pcc)$.
 \label{fig:davg}
 }
\end{figure*}

The evolution of S02 and S05 models is qualitatively similar to that
of S10 models in that phase transitions occur at the shock and in
the postshock expansion zone, although the former with low postshock
density do not produce much dense gas even under the uniform heating
rate. When the spiral potential is absent, the equilibrium disks of
these models consist entirely of the rarefied gas with the midplane
pressure $P(0)/\kb\sim3500-4000\pcc\Kel$ for S05 models and
$P(0)/\kb\sim1500-2000\pcc\Kel$ for S02 models. But, the shock
compression increases the pressure by about a factor of 3,
corresponding to a typical Mach number $\mathcal{M}\approx
v_{0x}/c_R\sim2$ for the rarefied medium,\footnote{ For adiabatic
shocks, the pressure jump condition is $P_{\rm 2}/P_{\rm
1}=1+\mathcal{M}^2(1-1/s)$, where the subscripts $1$ and $2$ denote
preshock and postshock values, respectively, and
$s=[(\gamma+1)\mathcal{M}^2]/[2\gamma+(\gamma-1)\mathcal{M}^2]$ is
the density shock jump factor.}  making the midplane postshock
pressure larger than $P_{\rm max}$. As a result, the dense medium in
S05 models comprises about 60\% of the total mass,  undergoing TI in
the postshock expansion zone. In S02 models, the postshock pressure
barely exceeds $P_{\rm max}$, so that the shocked dense gas,
comprising about 10\% of the total, easily disperses to return to
the rarefied gas in the interarm region. Flapping motions of spiral
shocks are correspondingly weaker in these models, with $\Delta
x/H_R\sim 0.7$ and $0.3$ for S05 and S02 models, respectively.

\subsection{Time-averaged Shock Structure}\label{sec:surf}
To visualize spiral shock structure in each model, we construct a
time-averaged distribution of number density $\langle n \rangle$.
Here, the angle brackets $\langle~\rangle$ denote a time average
over $t/\torb=5-8$. Figure~\ref{fig:davg} displays $\langle n
\rangle$ for all the non-self-gravitating models in logarithmic
color scale. It is apparent that S05 and S10 models possess a thin
dense layer everywhere near the midplane, while the dense gas is
found only inside the arm regions in S02 models. The shock
compression factors in the time-averaged density profiles are $\sim
7-10$, which is larger than the adiabatic shock jump due to enhanced
radiative cooling in the shocked gas \citep[cf.,][Paper
II]{muf74,ino08}. The shock transition layer in S05 and S10 models
is relatively broad because of rather strong flapping motions of the
shocks, while S02 models exhibit relatively sharp discontinuities.
Compared to NU models, arms in NM models have larger pressure and
are more expanded vertically, similar to ``hydraulic jumps'' that
occur when the equation of state is stiffer than isothermal (e.g.,
\citealt{mar98}). Table~\ref{tbl:arm} lists the time-averaged values
of the mass fractions and scale heights of dense and rarefied
components, as well as the overall average scale height.

\begin{figure}
 \epsscale{1.}
 \plotone{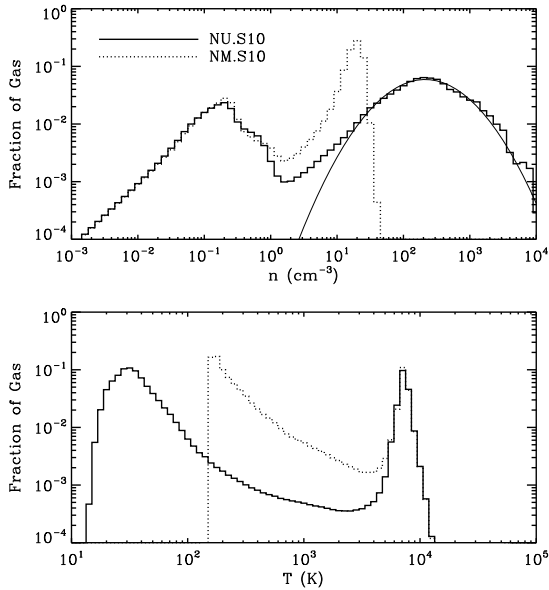}
 \caption{
 Mass-weighted density (\emph{top}) and temperature (\emph{bottom})
 probability density functions, averaged over $t/\torb=5$--8, in
 models NU.S10 (\emph{solid}) and NM.S10 (\emph{dotted}).
 While the broad dense peak centered at
 $(n,T)\sim(200\pcc, 30\Kel)$ in model NU.S10 is
 compressed and shifted to $\sim(20\pcc, 180\Kel)$ in model NM.S10, the
 rarefied peak at $(n,T)\sim(0.2\pcc, 7100\Kel)$ is unchanged.
 The thin line in the top panel is a lognormal fit to the dense peak in
 model NU.S10, with a standard deviation of $\Delta (\ln n) = 1.2$.
 \label{fig:pdf}
 }
\end{figure}

Figure~\ref{fig:pdf} plots the mass-weighted probability distribution
functions (PDFs), averaged over $t/\torb=5-8$,
of gas density and temperature for models NU.S10 and NM.S10.
The PDFs are in general bimodal, as is expected for a bistable cooling
function.
For model NU.S10, the dense and rarefied peaks are centered at
$(n, T)\sim(200\pcc, 30\Kel)$ and $\sim(0.2\pcc, 7100\Kel)$, respectively,
mostly distributed near the thermal equilibrium curves.
Only a small fraction of the gas is in the thermally-unstable regime.
The dense portion of the PDF in model NU.S10 is well fitted by a lognormal
distribution (\emph{thin solid line}) with a standard deviation of
$\Delta (\ln n) = 1.2$,
which is one of the characteristics of near-isothermal turbulence
\citep[cf.,][]{wad07,wad08}.
With enhanced heating, on the other hand, model NM.S10 shows a sharp
density cutoff in the density PDF at $n\sim 50\pcc$ and has a dense peak
shifted to $n\sim 20\pcc$.  Because of the stiff equation of state,
the dense gas in model NM.S10 is not as cold as in model NM.S10.
This not only thickens the midplane dense slab,
but also sets an upper limit on the gas density,
which in turn prevents the formation of dense condensations.
In model NM.S10, all the postshock dense gas becomes thermally unstable
in the expansion zone
and separates into dense and rarefied gas in the interarm region.

\begin{figure}
 \epsscale{1.}
 \plotone{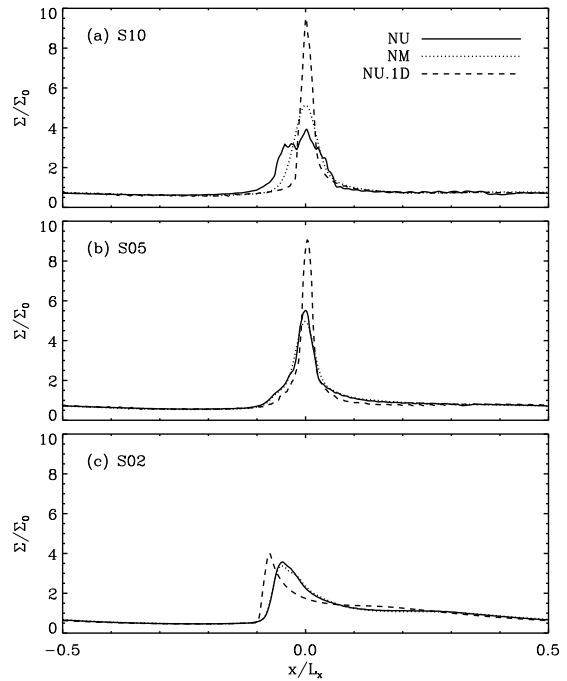}
 \caption{
 Time-averaged profiles of surface density from our two-dimensional
 simulations under the uniform heating rate (\emph{solid}) and
 the density-modified heating rate (\emph{dotted}) for
 models with (\emph{a}) $\Sigma_0=10\Surf$,
 (\emph{b}) $5\Surf$,
 and (\emph{c}) $2\Surf$.
 Dashed lines give the results of one-dimensional models
 (which do not have shock flapping)
 with the uniform heating rate.
 Stronger shock flapping motions in two-dimensional models make the arms broader
 and less peaked compared with the one-dimensional
 counterparts.
 \label{fig:surf}
 }
\end{figure}

Paper II showed that for one-dimensional models, the density
profiles of multiphase spiral shocks are more symmetric and have a
wider arms than isothermal counterparts. This is because the
strength of spiral shocks in the multiphase models fluctuates
depending on whether the incoming gas is warm or cold, resulting in
slight oscillations of the shock fronts in the direction
perpendicular to the arm. In addition, spiral shocks in the XZ plane
undergo flapping motions, which can further widen the arms. To see
this, we plot in Figure~\ref{fig:surf} the time-averaged surface
density profiles $\Sigma(x)=
\int_{-\infty}^{\infty}\langle\rho\rangle dz$ after taking a boxcar
average with window of $8\pc$. The solid and dotted lines are for NU
and NM models, respectively. Shown also as dashed lines are the
density profiles $n(x)/n_0$ from one-dimensional simulations (i.e.
without vertical stratification) under the uniform heating rate; the
initial number density $n_0$ of the one-dimensional counterpart was
chosen equal to the density-weighted mean density $n_{\rm
ave}=\Sigma_0 /(2\pi^{1/2}\mu m_{\rm H} H_{\rm ave})$, with the
average disk thickness $H_{\rm ave}$ listed in column (7) of
Table~\ref{tbl:arm}. For S05 and S10 models for which the shock
flapping motions are appreciable, the arms are considerably wider
and less centrally peaked than in the one-dimensional models. Due to
the flapping motions, dense condensates formed in the NU.S10 model
oscillate slightly in the $x$--direction when they pass through the
shock, resulting in broader arms than in the NM.S10 model. For S02
models, the one-dimensional shock consists only of the warm rarefied
gas, while the two-dimensional shocks contain a small amount ($\sim
10\%$) of dense gas due to an additional compression in the
$z$-direction; the difference in profiles is therefore slight.

\begin{figure}
 \epsscale{1.}
 \plotone{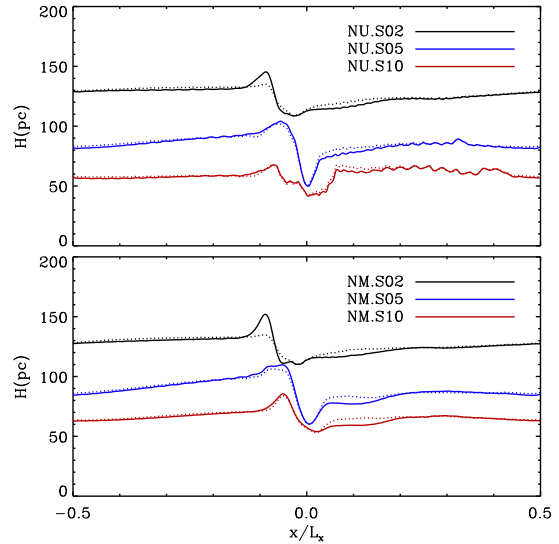}
 \caption{
 Vertical scale heights of the time-averaged density distributions in
 (\emph{a})  NU and (\emph{b}) NM models as functions of $x$.
 The simulation results (\emph{solid}) are overall in good agreement
 with the theoretical estimates (\emph{dotted})
 for effective vertical hydrostatic equilibrium.
 \label{fig:scaleh_nga}
 }
\end{figure}
\clearpage
\begin{deluxetable*}{lcccccccc}
\tabletypesize{\footnotesize}
\tablecaption{Induced Random Velocity Dispersions\label{tbl:veld}} \tablewidth{0pt}
\tablehead{
\colhead{} &
\colhead{} &
\multicolumn{3}{c}{Dense Component} &
\colhead{} &
\multicolumn{3}{c}{Rarefied Component} \\
\cline{3-5}\cline{7-9}
\colhead{Model} &
\colhead{$\Sigma_0$} &
\colhead{$\langle\delta v_x \rangle$} &
\colhead{$\langle\delta v_y \rangle$} &
\colhead{$\langle \delta v_z \rangle$} &
\colhead{} &
\colhead{$\langle \delta v_x \rangle$} &
\colhead{$\langle \delta v_y \rangle$} &
\colhead{$\langle \delta v_z \rangle$} \\
\colhead{   } &
\colhead{($\rm M_\odot\pc^{-2}$)} &
\colhead{(km s$^{-1}$)} &
\colhead{(km s$^{-1}$)} &
\colhead{(km s$^{-1}$)} &
\colhead{} &
\colhead{(km s$^{-1}$)} &
\colhead{(km s$^{-1}$)} &
\colhead{(km s$^{-1}$)}
} \startdata
         NU.S02 &      2 & $ 3.04 \pm 1.53$ & $ 3.28 \pm 1.23$ & $ 1.44 \pm 0.34$ &  & $ 2.62 \pm 0.58$ & $ 2.25 \pm 0.29$ & $ 1.61 \pm 0.17$ \\
         NU.S05 &      5 & $ 3.12 \pm 0.87$ & $ 3.20 \pm 0.92$ & $ 0.61 \pm 0.14$ &  & $ 3.04 \pm 0.34$ & $ 2.75 \pm 0.33$ & $ 1.64 \pm 0.14$ \\
         NU.S10 &     10 & $ 4.26 \pm 1.75$ & $ 3.60 \pm 1.89$ & $ 0.41 \pm 0.08$ &  & $ 3.47 \pm 0.81$ & $ 2.94 \pm 0.41$ & $ 1.72 \pm 0.14$ \\
         NM.S02 &      2 & $ 3.07 \pm 1.24$ & $ 3.24 \pm 1.20$ & $ 1.52 \pm 0.43$ &  & $ 2.79 \pm 0.40$ & $ 2.45 \pm 0.18$ & $ 1.82 \pm 0.11$ \\
         NM.S05 &      5 & $ 2.99 \pm 0.66$ & $ 3.20 \pm 0.81$ & $ 1.03 \pm 0.29$ &  & $ 3.17 \pm 0.59$ & $ 2.69 \pm 0.46$ & $ 1.74 \pm 0.11$ \\
         NM.S10 &     10 & $ 3.18 \pm 0.62$ & $ 3.51 \pm 1.08$ & $ 1.11 \pm 0.40$ &  & $ 3.07 \pm 0.42$ & $ 2.63 \pm 0.33$ & $ 1.69 \pm 0.12$ \\
         SM.S02 &      2 & $ 4.47 \pm 1.33$ & $ 3.97 \pm 1.49$ & $ 0.72 \pm 0.18$ &  & $ 3.04 \pm 0.50$ & $ 2.49 \pm 0.19$ & $ 1.85 \pm 0.14$ \\
         SM.S05 &      5 & $ 6.36 \pm 1.79$ & $ 6.95 \pm 2.11$ & $ 0.75 \pm 0.10$ &  &  $ 6.07 \pm 1.44$ & $ 5.50 \pm 1.45$ & $ 2.20 \pm 0.14$ \\
         SM.S10 &     10 & $10.52 \pm 5.55$ & $ 8.41 \pm 3.91$ & $ 3.60 \pm 1.71$ &  & $10.60 \pm 2.94$ & $ 7.92 \pm 1.78$ & $ 4.96 \pm 1.26$
\enddata

\tablecomments{Model name prefixes are as in Table \ref{tbl:noarm}.}

\end{deluxetable*}

In studies of galactic disk structure, it has been the customary to assume
effective hydrostatic equilibrium in the vertical direction.
Using numerical simulations without spiral arms, \citet{ko09b} explicitly
demonstrated that turbulent, multiphase disks are in effective
hydrostatic equilibrium, provided that the
turbulent pressure arising from random gas motions is taken into account.
When a spiral potential is present, the gas
surface density and velocity dispersions depend upon the distance $x$
from the minimum of the spiral potential.
It is interesting to study whether ``local'' effective hydrostatic
equilibrium is still established at each $x$.

From the time-averaged density distribution, we measure the density-weighted
vertical scale height $H(x)$, sound speed $c_s(x)$ , and
vertical velocity dispersion $\delta u_z(x)$ via
\begin{eqnarray}
H^2(x)=\frac{\int\langle\rho\rangle z^2dz}{\int\langle\rho\rangle dz},
\;\;\;\;
c_{s}^2(x)
=\frac{\int \langle P \rangle dz}
{\int\langle\rho\rangle dz},\nonumber\\
\;\;\;\;{\rm and}
\;\;\;\;
\delta u_{z}^2(x)
=\frac{\int \langle  \rho[v_z-\langle v_z\rangle]^2\rangle dz}
{\int\langle\rho\rangle dz}.
\end{eqnarray}
In the absence of self-gravity, the ``estimated'' vertical scale height
is given by
\begin{equation}\label{eq:hest}
H_{\rm est}^2(x)=\frac{c_s^2 + \delta u_{z}^2}{4\pi G\rho_*}
\end{equation}
for effective hydrostatic equilibrium \citep{ko09b}.
Figure~\ref{fig:scaleh_nga} plots $H(x)$ (\emph{solid lines})
and $H_{\rm est}(x)$ (\emph{dotted lines})
for NU and NM models as functions of $x$.
The measured vertical scale height is overall
in excellent agreement with the estimated value at all horizontal locations.
For all models, $c_s$ is about 5-7 times larger than $\delta u_z$.
This implies that the disks with spiral arms,
in time-averaged sense, are effectively in
\emph{vertical} hydrostatic equilibrium,
with the main support provided by thermal pressure (in these models
without stellar feedback).
Unsteady motions associated with shock flapping and movements of dense
condensations are mostly horizontal, affecting the vertical
force balance relatively little.
A small mismatch between $H$ and $H_{\rm est}$ near
$x/L_x=-0.1$ in S02 models arises from the fact that shocks in these models
exhibit weak flapping motions and retain sharp discontinuities in
the time-averaged configurations.
In this case, the
$\partial\langle \rho v_x v_z\rangle/\partial x$ term in the momentum
equation has a non-negligible contribution to the vertical force
balance, which was ignored in the derivation of equation (\ref{eq:hest}).

\subsection{Velocity Dispersions}\label{sec:turb}

Paper I showed that two-dimensional (isothermal) spiral shocks exhibit
strong flapping motions in the XZ plane and are able to generate a sonic
level of random gas motions in the arm regions.
On the other hand, Paper II showed that in one-dimensional spiral
shocks with TI, random gas motions amount to only $\sim1-2\kms$.
In this subsection, we quantify the level of random gas motions driven
by shock flapping motions and TI in our
two-dimensional models.

\begin{figure}
 \epsscale{1.}
 \plotone{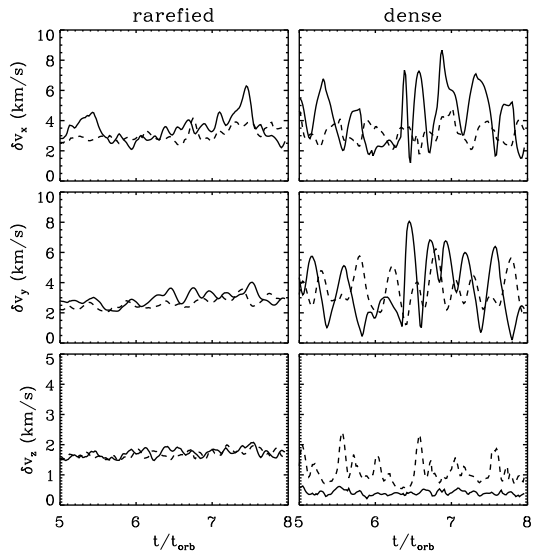}
 \caption{
  Temporal changes of the density-weighted velocity dispersions
  $\delta v_x$, $\delta v_y$, and $\delta v_z$
  of the rarefied (\emph{left}) and dense (\emph{right}) components in
  models NU.S10 (\emph{solid}) and NM.S10 (\emph{dashed}).
  The large-amplitude fluctuations
  of the velocity dispersions, with periods of $\sim 0.5\torb$,
  are due to incomplete subtraction of the arm streaming motions associated
  with the shock flapping.
 \label{fig:veld_tevol}
 }
\end{figure}

\begin{figure}
 \epsscale{1.}
 \plotone{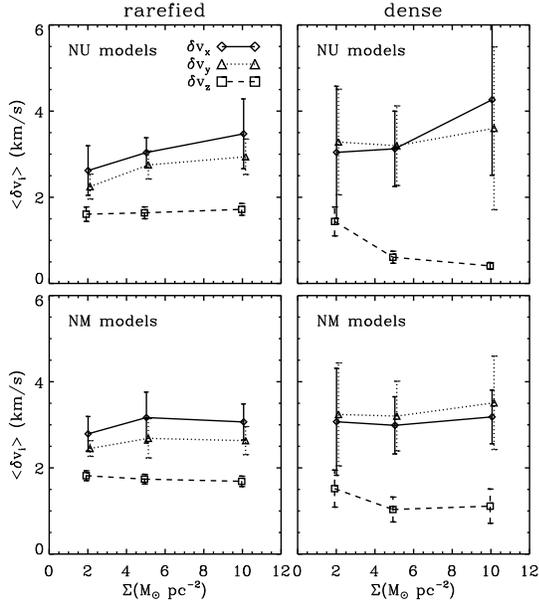}
 \caption{Mean values (\emph{symbols}) and
  standard deviations (\emph{errorbars}) of the density-weighted velocity
  dispersions, averaged over $t/\torb=5-8$, of the rarefied (\emph{left})
  and dense (\emph{right}) components in all non-self-gravitating models
  under the uniform heating rate (\emph{top}) and the density-modified heating
  rate (\emph{bottom}).  Allowing for the incomplete subtraction of the arm
  streaming motions, the random gas motions are
  $\sigma_x \sim \sigma_y \sim 2-3 \kms$ for both dense and rarefied
  components, and $\sigma_z \sim 1.7 \kms$ for the rarefied component,
  insensitive to $\Sigma_0$, in both NU and NM models, while
  $\sigma_z \propto \Sigma_0^{-0.8}$ for the dense component in
  NU models.
 \label{fig:veld}
 }
\end{figure}

The velocity field of gas moving across spiral arms is a combination of
several different components including streaming motions, oscillations
of the shock fronts themselves, and random motions.
Since streaming velocities that are ordered and vary with $x$ are much
larger than the true random motions of the gas,
it is important to subtract the former from the total velocity
as cleanly as possible.  For this purpose,
we first construct time-averaged templates of the
velocity field $\langle v_{i}\rangle$ (with $i=x$, $y$, or $z$)
for the dense and rarefied components separately.  We then calculate
the density-weighted velocity dispersions using
\begin{equation}
\delta v_{i}(t)=\left[\frac{\int \rho [v_{i}-\langle v_{i}\rangle]^2 dxdz}
{\int \rho dx dz}\right]^{1/2}.
\end{equation}
Figure~\ref{fig:veld_tevol} plots $\delta v_i(t)$ for the dense
and rarefied components in models NU.S10 and NM.S10
as solid and dashed curves, respectively, for a time span
of $t/\torb=5-8$.
Figure~\ref{fig:veld} plots the mean values $\vd$ along with
their standard deviations
$\Delta\delta{v}_{i}=(\langle \delta v_{i}^2 \rangle - \vd^2)^{1/2}$
for all the non-self-gravitating models; the values of
$\vd$ and $\Delta\delta{v}_{i}$ are listed in Table~\ref{tbl:veld}.

Figure~\ref{fig:veld_tevol} shows that the density-weighted velocity
dispersions for the dense component exhibit large-amplitude
fluctuations, with periods roughly of $\sim 0.5 \torb$.
The standard deviations of the fluctuations are
$\Delta\delta{v}_{i} \sim (0.2$--$0.5)\vd$ for the dense component;
deviations are only
$\Delta\delta{v}_{i} \sim (0.1$--$0.2)\vd$ for the rarefied phase.
These variations of $\delta v_x$ and $\delta v_y$ are caused mostly by
oscillations of the shock front relative to the mean position.
With large spatial variations of streaming velocities across the
arm, the small offset of the shock position as well as the
instantaneous locations of dense condensates result in large values
of $\Delta\delta{v}_{i}$. We thus regard the local minima of
$\delta{v}_i$, approximately equal to $\sigma_i \equiv
\vd-\Delta\delta{v}_i$, as the upper limits to the level of random
gas motions.

Figure~\ref{fig:veld} shows that for NU models, $\langle\delta{v}_x\rangle$
and $\langle\delta{v}_y\rangle$ increase with $\Sigma_0$.
This is mainly because the shock compression and associated phase transition
are stronger with larger $\Sigma_0$, leading to stronger flapping motions.
Nevertheless, $\sigma_x \sim \sigma_y \sim 2-3\kms$ for both dense and
rarefied components, insensitive to $\Sigma_0$.  This indicates that
the portion of kinetic energy in the shock flapping motions that
goes into random gas motions is quite limited.  The remaining portion
is simply associated with the horizontal shock oscillations near the
midplane.
Since the shock flapping motions at low $|z|$ are mostly
horizontal, the random vertical motions of the dense gas
in NU models are forced predominantly by the impact of rarefied gas
arriving from high altitudes.  The fact that the rarefied gas
has $\sigma_z\sim 1.7\kms$, almost independent of $\Sigma_0$, suggests
the flapping motions drives  more-or-less constant vertical motions
at high $|z|$; this is because the total mass of rarefied gas is
almost the same in all models, equivalent to a surface density of $1.9
\Msun \pc^{-2}$.
Since the fraction of the rarefied component decreases with increasing
$\Sigma_0$ (see Table~\ref{tbl:arm}), the ratio of vertical kinetic energy
in the rarefied gas
to the mass of dense gas
decreases with $\Sigma_0$.  This causes
$\sigma_z$ of the dense medium to decrease
with increasing $\Sigma_0$, roughly as
$\sigma_z \propto \Sigma_0^{-0.8}$ in NU models.
For NM models, the dense gas in the immediate postshock region
is overpressured due to the strong heating and thus expands vertically,
enhancing $\sigma_z$ compared to those in NU models.

\section{Self-Gravitating Models}

\begin{figure*}
 \epsscale{1.}
 \plotone{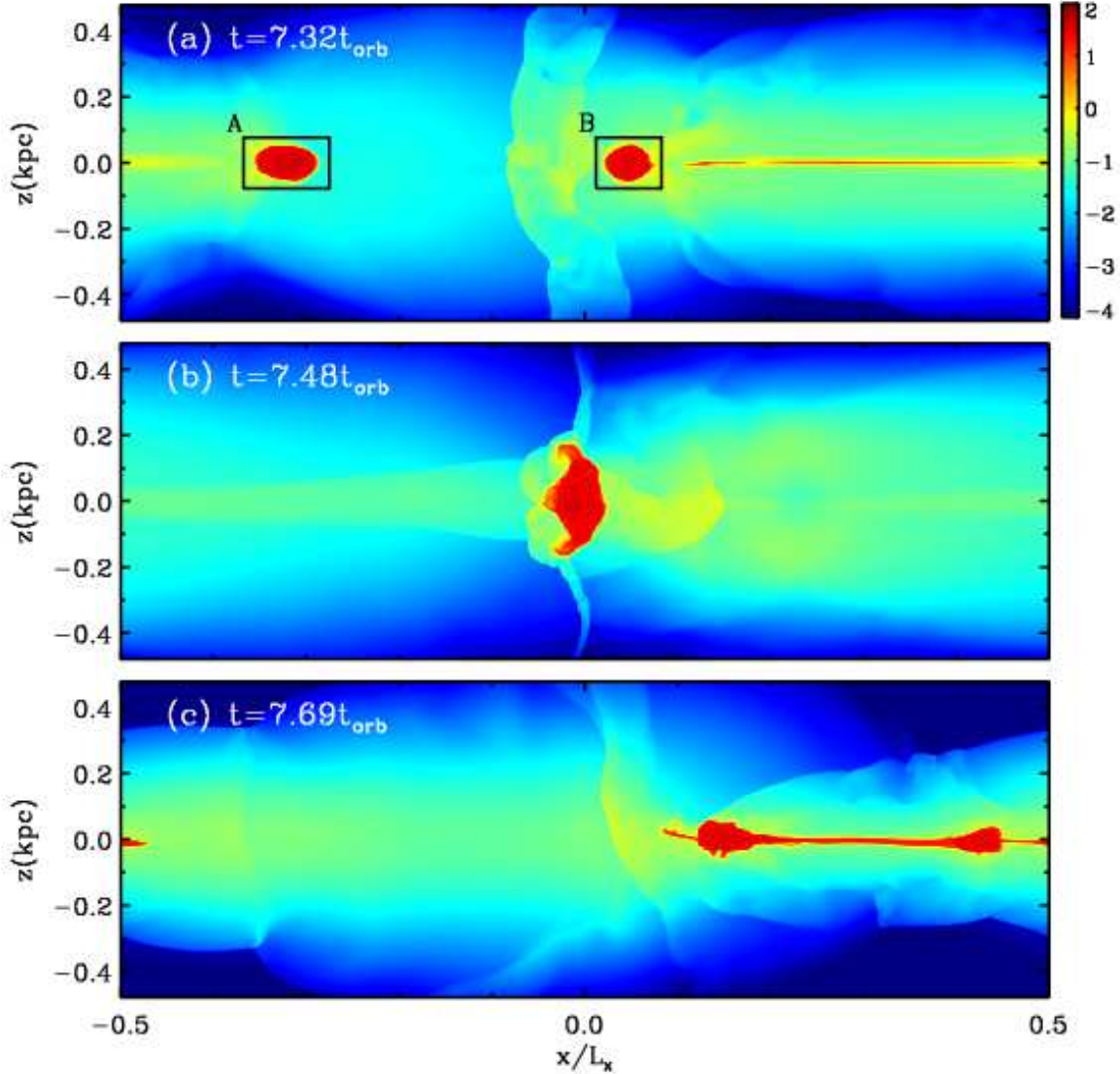}
 \caption{Density snapshots of self-gravitating model SM.S10
  at $t/\torb=7.32$, 7.48, and 7.69 in logarithmic color scale.
  Two clouds are separate from each other in the interarm regions
  (\emph{a}), temporarily merge together at the shock (\emph{b}),
  and then break up back into two pieces in the postshock expansion zone
  (\emph{c}).
  The boxes surrounding the clouds A and B in (\emph{a}) are enlarged
  in Figure~\ref{fig:clump} to show the interval velocity structure.
  Colorbar labels $\log (n/1\pcc)$.
  \label{fig:sgb}
 }
\end{figure*}

We now explore the formation of
self-gravitating clumps and their internal properties.
For  NU models,
we find that the inclusion of self-gravity always results in catastrophic
collapses of self-gravitating clouds that form in the postshock
region, preventing us from continuing simulations further.
For this reason, we present the results of self-gravitating models
only with the density-modified heating rate (``SM'' models).
Instead of running self-gravitating models from $t=0$,
we make use of the ``saturated-state''
data sets of NM models at
$t/\torb\sim4.8$ and restart them by slowly turning on the
gaseous self-gravity over a time interval of $1.5\torb$.

Neglecting the effect of the rarefied medium, the gravitational
susceptibility of a midplane dense layer in
NM models can be measured by the average Toomre stability parameter
\begin{equation}
Q_D = \frac{1}{f_D}\frac{\kappa c_D}{\pi G \Sigma_0}
= \frac{0.27}{f_D} \left(\frac{\Sigma_0}{10\Surf}\right)^{-1},
\end{equation}
where $c_D=1\kms$ is the mean sound speed of the dense gas. With the
inclusion of self-gravity, the dense layer in model SM.S10 has
$Q_D=0.34$ and thus is quite unstable, initiating the collapse of
high-density regions. Due to the stiff equation of state, however,
the collapsing clouds soon reach an equilibrium state with only a
moderate central density $30 \pcc$, which is only $1.5$ times larger
than the average density of the dense gas in model NM.S10. As these
clouds follow galaxy rotation, they merge together in regions of
converging streaming velocities, eventually resulting in two big
condensations. Figure~\ref{fig:sgb} shows the density snapshots in
logarithmic color scale of model SM.S10 at $t/\torb=7.32$, $7.48$,
and $7.69$. Two clouds are widely separated from each other during
traversal of the interarm region (Fig.~\ref{fig:sgb}\emph{a}). After
one cloud enters the spiral shock, it loses most of its
$x$-momentum, and the second cloud can then collide with it when it
enters the shock at high speed (Fig.~\ref{fig:sgb}\emph{b}). Since
the two clouds are on their own epicyclic orbits before the
collision, however, they retain quite different $v_y$ even after the
collision, preventing them from merging into a single entity. Due to
the Coriolis force, the two clouds subsequently have different
$v_x$, so that the merged entity elongates in the expansion zone
(Fig.~\ref{fig:sgb}\emph{c}), and separates back into two pieces in
the interarm region.

The dense layer in model SM.S05 has $Q_D=0.9$ and is thus marginally
unstable.  With the aid of shock compression, the dense gas in the
postshock region collapses and eventually forms two self-gravitating
clumps. In this model, the collision of these clumps at the shock
and subsequent break up in the expansion zone is similar to model
SM.S10. With $Q_D=12$, on the other hand, the dense layer in model
SM.S02 is gravitationally stable and does not form dense clumps.
Compared with model NM.S02, self-gravity in model SM.S02 increases
the fraction of the dense phase by about a factor of 2, which in
turn decreases its vertical velocity dispersion by a similar factor.

The presence of self-gravity leads to stronger shock flapping
motions than in the NU models, increasing the velocity dispersion.
In model SM.S02, the in-plane velocity dispersions of the dense
component increases by a factor of $\sim 1.2-1.5$ compared to model
NU.S02; for low surface density, the self-gravity is not
sufficiently strong to have a major effect.  For the SM.S05 and
SM.S10 models, however, the stronger self-gravity make a larger
difference to shock flapping, in turn driving larger velocity
dispersions. Correcting for streaming, the velocity dispersions of
both dense and rarefied phases in model SM.S05 reach
$\sigma_x\sim\sigma_y\sim4-5\kms$, about twice larger than those in
model NM.S05. In model SM.S10, the dense gas velocity dispersions
are $\sigma_x\sim\sigma_y\sim 4-5\kms$, while the rarefied gas has
the velocity dispersions up to $\sim 7\kms$.  Note that these
in-plane velocity dispersions in multi-phase, self-gravitating models
are similar to those in the isothermal self-gravitating
models with $F=5\%$ studied in Paper I.

\begin{figure*}
 \epsscale{1.}
 \plotone{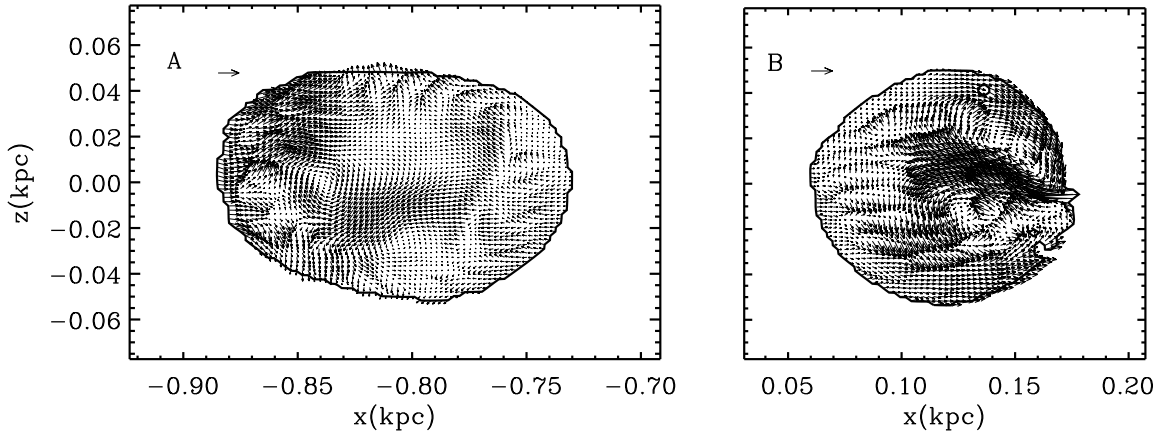}
 \caption{
 Shapes and random velocity fields of the clouds A and B shown in
 Figure~\ref{fig:sgb}\emph{a} for model SM.S10.
 The velocity vectors are distributed quite randomly, indicative of
 (subsonic) turbulent motions within the clouds.
 The arrow outside the clump in each panel
 corresponds to the velocity of $5\kms$.
 The clouds are gravitationally bound with virial parameter of
 $\alpha \sim 2$.
 \label{fig:clump}
 }
\end{figure*}

\begin{deluxetable*}{lcccccc}
\tabletypesize{\footnotesize}
\tablecaption{Average Properties of Clumps Produced in Self-Gravitating Models
\label{tbl:clump}} \tablewidth{0pt}
\tablehead{
\colhead{Model} &
\colhead{$c_{\rm cl}$} &
\colhead{$\sigma_{\rm cl}$} &
\colhead{$R_{\rm cl}$} &
\colhead{$n_{\rm cl}$} &
\colhead{$M_{\rm cl}$} &
\colhead{$\alpha$} \\
\colhead{     } &
\colhead{(km $s^{-1}$)} &
\colhead{(km $s^{-1}$)} &
\colhead{(pc)} &
\colhead{(cm$^{-3}$)} &
\colhead{$(10^6\Msun)$} &
\colhead{}
}
\startdata

SM.S05 & $ 3.69 \pm 0.67$ & $ 0.96 \pm 0.24$ & $ 44.9 \pm  7.6$ &
$ 27.4 \pm  0.8$ & $0.36 \pm 0.15$ & $2.34 \pm 0.25$ \\
SM.S10 & $ 5.06 \pm 0.63$ & $ 1.53 \pm 0.61$ & $ 60.7 \pm  6.0$ &
$ 29.8 \pm  1.6$ & $0.91 \pm 0.29$ & $2.31 \pm 0.32$
\enddata
\end{deluxetable*}

Since the self-gravitating clumps produced in models SM.S05 and SM.S10 move
almost ballistically, the position of the largest surface density at a
given time does not always correspond to the minimum of the spiral
potential.  In fact, the clumps are near the
potential minimum ($|x|/L_x < 0.1$) only for $\Delta t/\torb=0.35$,
making the definition of the arm regions rather ambiguous.
In addition, due to accretion onto the clumps, the rarefied medium in these
models amounts to less than 10\% of the total mass, much smaller than the observed
mass fraction of the warm gas near the solar neighborhood \citep{hei01,hei03}.
For these reasons, these clumps are unlike real
self-gravitating clouds in spiral galaxies.
Nevertheless, we believe these \emph{model  clouds} may provide clues
to the internal
properties and virial balance of real interstellar clouds, in that
they represent a limiting situation in which internal turbulent
feedback from star formation is absent.

Keeping in mind the caveats mentioned above, we proceed as follows to
calculate the cloud properties.  To define the boundary of the dense clumps,
we choose the threshold density $n_{\rm th}=21\pcc$, corresponding
to $P_{\rm max}$ in the thermal equilibrium curve.
Using a CLUMPFIND algorithm (e.g., \citealt{wil94}), we find the interior
of each cloud with $n>n_{\rm th}$.
We then measure the mean density $\rho_{\rm cl}\equiv \overline{\rho}$,
the averaged sound speed $c_{\rm cl}\equiv (\overline{P/\rho})^{1/2}$
and the mean one-dimensional internal velocity dispersion
$\sigma_{\rm cl}\equiv \sum_i (\overline{v_{i}^2}-\overline{v_{i}}^2)^{1/2}
/3^{1/2}$ of each cloud,
where the overlines denote the mass-weighted average.
We then count the total number of pixels $N_{\rm cl}$ on the XZ plane
occupied by each cloud, and calculate the cloud size
$R_{\rm cl}\equiv (N_{\rm cl}\Delta x \Delta z /\pi)^{1/2}$.
Assuming a spherical shape, we calculate the total mass
$M_{\rm cl}\equiv 4\pi\rho_{\rm cl} R_{\rm cl}^3/3$ and
the virial ratio of each cloud via
\begin{equation}
\alpha\equiv\frac{5(\sigma_{\rm cl}^2+c_{\rm cl}^2)R_{\rm cl}}{GM_{\rm cl}},
\end{equation}
which is the ratio of the total kinetic energy to the gravitational potential
energy for a uniform spherical cloud (e.g., \citealt{ber92,mo07});
note that central concentration would decrease $\alpha$.
Figure~\ref{fig:clump} gives an example of the shapes and internal velocities
of two clumps shown in Figure~\ref{fig:sgb}\emph{a}.
Note that the velocity vectors are distributed quite randomly,
indicative of (subsonic) turbulent motions within the clouds.

Table~\ref{tbl:clump} summarizes the average properties of the clouds
that form in models SM.S05 and SM.S10.
The typical size and mass of the clouds are found to be
$R_{\rm cl}\sim45$-$60\pc$ and $M_{\rm cl}\sim (4-9) \times 10^5\Msun$,
respectively, with the clouds in model SM.S05
somewhat smaller and less massive than in model SM.S10.
Overall, $\alpha\sim 2$ for all the clouds,
suggesting that they are (marginally) gravitationally bound.
The mean sound speed inside the clouds is
$c_{\rm cl}=3.7-5.1\kms$, $\sim 3-4$ times larger than
the one-dimensional internal velocity dispersions
$\sigma_{\rm cl}=0.9-1.8\kms$.  This indicates that
the major support against self-gravity comes predominantly
from the thermal energy,  a consequence of the
density-modified heating rate we adopt.  The relatively low value of
the internal turbulent velocity dispersion suggests that the
interaction of a dense, gravitationally-bound cloud with its surroundings
can drive only a modest level of internal turbulence.

\section{Summary \& Discussion}

While stellar spiral arms in disk galaxies provide smoothly varying
low-amplitude gravitational potential perturbations, the response of
the interstellar gas to them is quite dramatic. Spiral shocks
compress the ISM and the high post-shock densities  may trigger
growth of arm substructures and star formation. In addition,
radiative cooling and heating of the gas makes the ISM inherently
inhomogeneous, producing two phases that differ in density and
temperature by about two orders of magnitude. Moreover, the vertical
stellar gravity tends to produce stratification of the cold and warm
gas due to differential buoyancy; this stratification can be
modified by vertical turbulence, however. Interactions of these
processes may significantly affect the gas flows and shock
structures, compared with results from our previous work (and that
of other groups), which employed an isothermal approximation (Paper
I) or neglected the vertical degree of freedom (Paper II). In this
paper, we have conducted nonlinear hydrodynamic simulations in a
two-dimensional slice perpendicular to a local segment of a spiral
arm that is tightly wound (with a pitch angle
$\sin i =0.1$) and rotates rigidly (with a pattern speed
$\Omega_p/\Omega_0=0.5$). To handle the Coriolis force arising from
the galaxy rotation self-consistently, we allow for gas motions
parallel to the arm (i.e. perpendicular to the domain of the
simulation). We consider two different forms of gas heating; the
usual constant heating rate (for NU models) and the density-modified
heating rate (for NM and SM models), which mimics the effect of
star-formation feedback in a very simple way, to limit runaway
collapse. We start from initially isothermal disks that are in
vertical hydrostatic equilibrium but out of thermal equilibrium. We
slowly turn on the amplitude of the spiral arm potential such that
it attains a full strength at 1.5 orbital times. Magnetic fields are
neglected in the present work.

Our main results and their astronomical implications are as follows.

1. \emph{Two-phase disk equilibria without spiral arms.} ---
In the absence of spiral-arm potential perturbations (and other
sources of turbulence), the vertical
structure of equilibrium disks
depends on the disk surface density $\Sigma_0$.
When $\Sigma_0>\Sigma_{\rm max}\equiv P_{\rm max}/(2G\rho_* c_R^2)^{1/2}
\rightarrow 6.7\Surf$ (for Solar neighborhood conditions),
the disk experiences TI and evolves toward an equilibrium
configuration with a thin slab of dense gas ($n>1\pcc$)
near the midplane
sandwiched between layers of rarefied gas (with $n<1\pcc$).
Here, $P_{\rm max}/\kb=5000\pcc\Kel$ is the maximum pressure allowed
for the thermally-stable rarefied gas with our adopted heating and
cooling functions, $c_R\approx 7\kms$ is its
density-weighted sound speed, and $\rho_*=0.056\Msun\pc^{-3}$ is the
stellar density near the Solar neighborhood.
In our models, the transition between the dense and rarefied phases
occurs approximately at $P_{\rm trans}/\kb\approx 2100\pcc\Kel$,
insensitive to $\Sigma_0$, as long as the disk is unstable to TI.
Without self-gravity, the vertical distribution of the rarefied gas
in equilibrium is well fitted by a Gaussian profile whose surface
density is fixed to $\Sigma_R = \SigNG \equiv P_{\rm
trans}/(2G\rho_*c_R^2)^{1/2} \rightarrow 2.8\Surf$.
When self-gravity is included, the
gravity from the midplane dense layer compresses the overlying
rarefied component further, forcing overpressured rarefied gas to
transform to the dense phase. The resulting surface density of the
rarefied gas is reduced to $\Sigma_R ={\mathcal F}\SigNG$, with the
reduction factor ${\mathcal F}$ defined by equation (\ref{eq:reduc}).
When $\Sigma_0<\Sigma_{\rm min}\equiv P_{\rm min}/(2G\rho_*c_R^2)^{1/2}
\rightarrow 2.1\Surf$, the disk has too low a pressure
to produce the dense component; equilibrium consists only of the
rarefied gas. When $\Sigma_{\rm min}<\Sigma_0<\Sigma_{\rm max}$,
either a single rarefied disk or a two-phase disk is possible,
depending on the initial conditions. Our S02 and S05 models with
$\Sigma_0=2$ and $5\Surf$, respectively, that start from a warm
isothermal configurations all end up with a single-component
rarefied disk, in the absence of spiral perturbations.

2.  \emph{Shock flapping motions in vertically stratified disks.} ---
Spiral-arm potential perturbations
lead to spiral shocks in the gas, which are
vertically curved and non-stationary,
showing strong flapping motions
perpendicular to the arms.
Similarly to the one-dimensional cases studied in Paper II,
the shock compression and postshock expansion in two dimensions
allow phase transitions, but only if the gas density at the shock
and/or the postshock expansion zone reaches the thermally-unstable range
($1\pcc \simlt n \simlt 7-9\pcc$).
In model NU.S10 with $\Sigma_0=10\Surf$ and
the uniform heating rate, the shocked dense gas has large enough  density
that the postshock expansion is unable to return it to the thermally
unstable regime.
As a consequence, there is a large amount of interarm dense gas
entering the shock in this model, which collides with other dense gas
in the arm, producing dense condensations.
In other models with lower surface density or the density-modified heating
rate, the shocked gas re-expands and
becomes thermally unstable,
returning to either the dense or the rarefied phase in the interarm region.

The shock flapping motions in our models arise due to the fact that
the arm crossing time of gas is incommensurate with the vertical
oscillation period, so that steady flows are not possible. Seen from
the upstream side, the shock is convex when the postshock regions
are maximally compressed, and concave when the postshock vertical
expansion is strongest. These non-steady motions of shock fronts are
commonly seen in numerical simulations with sufficient resolution
(e.g., \citealt{mar98, gom02, gom04, kim06}; Paper I;
\citealt{wad08}), although simulations with low resolution
\citep{tub80} or particles (e.g., \citealt{dob08c}) do not capture
the flapping motions clearly.

It is interesting to note that radio continuum images
of the 5-kpc arm (or the Scutum arm) toward the galactic
longitude $l=30^\circ-32^\circ$ in the Milky Way shows
a bow shock feature in the warm ionized medium
with temperature $\sim10^4\Kel$ \citep{sof85}, similarly to
a convex shock front seen in Figure~\ref{fig:na10}\emph{b}.
The radio emission from the bow shock is presumably thermal
radiations from ionized gas, with emission measure
of $\sim 7,000\pc$ cm$^{-6}$.
The curvature of the observed bow shock, as measured by
the longitudinal offset $\Delta l$ of the shock at latitude $b$
relative to the shock front at the midplane,
is $\Delta l/b\sim 0.5$ for $b=0.5^\circ$.
This value is about a half of the maximum curvature of
the shock front $|x_{\rm sp}(H_R)-x_{\rm sp}(0)|/H_R\sim0.85$
in our S10 models, where
$x_{\rm sp}(z)$ denotes the shock position at height $z$.
This strongly suggests that the bow shock associated with the 5-kpc arm is
most likely a cross section of a galactic spiral shock that
is undergoing flapping motions.
Velocity information is needed to determine whether
the 5-kpc arm regions are currently being compressed or expanding in
the course of the flapping motions.

In this paper, for consistency with our local
approximation we have considered tightly wound arms with a
very small pitch angle $i\sim5.8^\circ$, and a pattern speed
$\Omega_p/\Omega_0=0.5$. Observed spiral arms of external galaxies
are often more loosely wound with $i\sim10^\circ-30^\circ$
\citep[e.g.,][]{sei08} and span a wide range of galactocentric radii
with differing $\Omega_p/\Omega_0$.  For fixed $F$, a larger arm pitch
angle would imply a larger streaming velocity $v_{\rm 0x}$
perpendicular to the arm (see eq.\ \ref{eq:vel0}). Spiral shocks would
then become correspondingly stronger and shifted farther downstream
\citep[e.g.,][]{rob69,shu73,kim02}, exhibiting larger amplitude
flapping motions (Paper I). On the other hand, $|v_{\rm 0x}|$ is
increasingly small as $\Omega_p$ approaches $\Omega_0$.  Consequently,
the spiral shock as well as associated flapping motions would become
weaker as one approaches corotation, where the gas
would simply concentrate near the potential minimum,
without involving a shock.

3. \emph{Time-averaged shock structure.} --- Within a few orbital
times after the development of spiral shocks, gas flows reach a
quasi-steady state in the sense that the mass fractions of dense and
rarefied gas do not change appreciably with time. For models with
$\Sigma_0=10\Surf$, the quasi-steady mass fraction of the rarefied
gas is $f_R \sim 19\%$, which can be compared to $f_R\sim30\%$ when
the spiral potential is absent.  Despite the shock flapping motions,
most of the gas is found close to thermal equilibrium, with a small
fraction thermally unstable. The density and temperature PDFs are
thus bimodal. For model NU.S10, the dense and rarefied peaks are
located  at $(n, T)\sim(200\pcc, 30\Kel)$ and $\sim(0.2\pcc,
7100\Kel)$, respectively, and the dense part of the density PDF is
described by a lognormal distribution. The time-averaged structure
can be well represented by \emph{local} vertical hydrostatic
equilibrium, supported mainly by the thermal pressure rather than
gas random motions. This indicates that the vertical hydrostatic
balance is a reasonable approximation even in the presence of spiral
shocks. The profiles of surface density perpendicular to the arm are
more-or-less symmetric with a shock compression factor of
$\sim 7-10$, and have broad arm regions whose width correlates with
the strength of the shock flapping motions. The fractional
widths of the arm, postshock expansion zone, and interarm region are
typically 10\%, 20\%, and 70\% of the arm-to-arm distance, where the
gas stays for 15\%, 30\%, and 55\% of the arm-to-arm crossing time,
respectively. The shock flapping motions in the XZ plane make the
arm wider than in one-dimensional spiral shocks where the arm takes
up only 1\% of the arm-to-arm distance (Paper II).

The dense gas produced from TI and shock compression tends to sink
toward the midplane to form a thin slab, while high-altitude regions
are dominated by warm rarefied gas.  The thickness of the dense slab
is $H_D\sim 10-40\pc$, depending on the total gas content, heating
rate, and presence/absence of self-gravity, while the scale height
of the rarefied gas is $H_R\sim 130\pc \approx c_R/\sqrt{4 \pi G
\rho_*}$ insensitive to the parameters. For model NM.S10, the
thickness ratio of the dense to rarefied components is about 5,
which is not much different from the results of \citet{dob08c} who
reported that the warm gas extends vertically up four times more
than the cold gas. With high density and pressure, the dense slab
would transform to molecular clouds if the appropriate chemical
reactions for molecule formation were included (e.g.,
\citealt{dob07,dob08b}). Thin distributions of the cold dense gas
are in fact common in numerical simulations of galactic disks with
TI where turbulence is driven by magnetorotational instability
\citep{pio07}, stellar feedback via \ion{H}{2} regions
\citep{ko09a,ko09b}, or supernovae explosions
\citep{kor99,avi01,jou06,jou09}. The observed molecular distribution
the Milky Way has a scale height of $\sim 35\pc$ within the Solar
circle, somewhat reduced for the most massive clouds (e.g.,
\citealt{mal94,bro00,sta05}). Galactic \ion{H}{2} regions are also
within about $30\pc$ of the midplane \citep{loc77}. In the inner
Galaxy and near the Solar circle, the scale height of the cold
\ion{H}{1} layer is about 1.5 times smaller than the warm \ion{H}{1}
gas (e.g., \citealt{fal73, cro78}; see also \citealt{fer01}),
although the cold and warm phases appear to have a similar scale
height in the outer Galaxy (e.g., \citealt{dic09}).

4. \emph{Random gas motions driven by shock flapping motions.}  ---
The flapping motions of spiral shocks stir the gas and supply random
kinetic energy. Allowing for incomplete subtraction of streaming
motions in the arm region, the induced density-weighted velocity
dispersions are $\sigma_x\sim\sigma_y \sim 2-3\kms$ for both dense
and rarefied components for the non-self-gravitating models, with
larger values corresponding to disks with larger $\Sigma_0$.
Compared with the results of Paper II where the vertical coordinate
was suppressed, these values are similar to those for the dense gas
in the arms and larger by a factor of $\sim2-3$ for the interarm
rarefied gas.  This implies that it is the rarefied gas that is more
efficiently stirred by the shock flapping motions.  The
self-gravitating models have larger velocity dispersions, in the
range  $\sigma_x\sim\sigma_y \sim 4-5\kms$ for the dense and
$\sigma_x\sim\sigma_y \sim 4-7\kms$ for the rarefied gas, indicating
that self-gravity enhances shock flapping and velocity dispersions,
especially for rarefied gas. These in-plane velocity dispersions in
the current multiphase models are similar to those in the isothermal
models considered in Paper I.

The vertical velocity dispersions of the rarefied gas in NU and NM
models are $\sigma_z\sim 1.7\kms$, insensitive to $\Sigma_0$. In NU
models, the vertical motions of the dense gas are excited
preferentially by vertical motions of the rarefied gas. Since the
mass fraction of the rarefied gas decreases with $\Sigma_0$, the
vertical velocity dispersions of the dense gas in NU models is a
decreasing function of $\Sigma_0$, varying roughly as $\sigma_z
\propto \Sigma_0^{-0.8}$. In NM models, the postshock gas is
overpressured due to enhanced heating and thus expands vertically,
increasing $\sigma_z$ compared to NU models.

The level of random gas motions in our models are generally smaller
than the observed velocity dispersions $\sim 7-10\kms$ for atomic
gas in the solar neighborhood (e.g., \citealt{hei03}) and for the
larger molecular clouds in the Milky Way (e.g.,
\citealt{sta89,sol87,hey09}).  Thus, we conclude that other sources
of the interstellar turbulence (e.g., \citealt{mac04,elm04}) must
exceed that provided by spiral shocks. One of the dominant
mechanisms is of course supernova explosions
\citep[e.g.,][]{kor99,avi05,jou06,jou09}. In outer regions of
galaxies where star formation activity is low, the magnetorotational
instability \citep[e.g.,][]{sel99,pio05,pio07} and/or cosmic infall
of gas (e.g., \citealt{san07,san09,kle09}) may play an important
role in driving the ISM turbulence. \ion{H}{2} region expansion and
radiation pressure are important in injecting energy into the ISM as
GMCs are dispersed (e.g., \citealt{mat02,mur10}). At large scales,
self-gravitating instability with galactic rotation and shear can
drive turbulence at near-sonic levels
\citep[e.g.,][]{kim02,wad02,kim07,age09}.

5. \emph{Effect of self-gravity and properties of self-gravitating clouds.} ---
When self-gravity is included, dense gas in SM models with
$\Sigma_0 \geq 5 \Surf$ suffers from gravitational instability,
eventually forming two large clouds in each model.
These are separate in the interarm regions,
temporarily merge in the arm,
and then break up into two pieces in the postshock expansion zone.
These clouds have a radii
$\sim45-60\pc$ and mass $\sim(4-9)\times 10^5\Msun$ each, and are
gravitationally bound with a virial parameter of $\alpha\sim2$.
In our present models, we have not attempted to include realistic
star formation feedback, but instead increase the heating rate at high
density to prevent collapse.  As a consequence,  the main support
against self-gravity comes from thermal pressure.  The mean
thermal sound speed and internal velocity dispersion of the clouds
are $c_{\rm cl}\sim 3.7-5.1 \kms$ and $\sigma_{\rm cl}\sim 0.9-1.8\kms$,
respectively.
For models with $\Sigma_0=2\Surf$, self-gravity is insufficient to
form bound clouds.  Nevertheless,
self-gravity increases the dense gas fraction
by a factor of $\sim 2$ compared to the non-self-gravitating
counterpart of this model,
which in turn decreases the vertical velocity dispersion of the dense gas
by a similar factor.

Formation of self-gravitating clouds in our two-dimensional models
requires the production of the dense gas due to TI, and then
additional shock compression. Although our present models do not
capture the cloud destruction process, bound clouds created inside
spiral arms may be disrupted before they leave the arms if feedback
from star formation is sufficiently strong \citep{she08, wad08}.
Nevertheless, the presence of high-density, self-gravitating
clouds in the interarm regions opens an interesting possibility that
the spiral shocks -- where the diffuse gas is strongly compressed --
do not necessarily coincide with the regions of highest
gas density (in gravitationally-bound clouds). For example,
\citet{pat06} found strongly-polarized nonthermal radio
emission that may trace magnetic arms, detected preferentially
upstream of the CO arms in the inner disk of the Whirlpool galaxy
M51 (see also e.g., \citealt{fle10}). We note, however, that the
current unmagnetized models are not yet able to
provide clues to the relation between gaseous and magnetic
arms. It will be interesting to see how TI, spiral shocks, and
realistic star formation feedback conspire with magnetized
self-gravitating instabilities to create bound clouds and arm
substructures (possibly including separate magnetic arms),
and to generate turbulence in the gas.

\acknowledgments The authors are grateful to B.-C.\ Koo and G.\ Park
for information on Galactic spiral arms, and also to the
referee for an insightful report. This work was supported in part
by the Korea Research Foundation Grant funded by the Koran
Government (MOEHRD) (KRF -- 2007 -- 313 -- C00328) and in part by by
KICOS through the grant K20702020016-07E0200-01610 provided by MOST.
Simulations were performed by using the supercomputing resource of
the Korea Institute of Science and Technology Information through
the grant KSC-2009-S02-0008. The work of ECO is supported by the
U.S. National Science foundation under grant AST-0908185.

\appendix

\section{Two-Phase Disks in Hydrostatic Equilibrium}

We consider thermally bistable two-phase disks in which a cold,
dense layer with surface density $\Sigma_D$ with thickness $H_D$ is
surrounded by a warm, rarefied medium. Since the scale height of the
dense layer is very small compared to that of the rarefied gas, we
approximate the former as razor-thin ($H_D\approx 0$).
We further assume that the mass
fraction of the rarefied gas is small, so that its self-gravity is
unimportant. Let $\rho_R(z)$ denote the density distribution of the rarefied
gas. In the presence of the external gravity from a stellar disk of uniform
density $\rho_*$, the condition of vertical hydrostatic equilibrium
for the warm gas reads
\begin{equation}\label{eq:hser}
c_R^2\frac{d \ln\rho_R}{d z}= -4\pi G\rho_*z-2\pi G \Sigma_D {\rm
sign}(z),
\end{equation}
where $c_R$ is the isothermal sound speed of the rarefied gas, assumed
to be independent of $z$. Integrating equation~(\ref{eq:hser}) over
$z$, one obtains
\begin{equation}\label{eq:rhor}
\rho_R(z)=\rho_R(0)\exp\left[-\frac{1}{2h_g^2}\left(z^2+\frac{\Sigma_D}{\rho_*}|z|\right)\right]
\end{equation}
where $h_g^2=c_R^2/(4\pi G\rho_*)$ is the Gaussian scale height
which the rarefied gas would have in the absence of self-gravity.
The surface density, $\Sigma_R$, of the rarefied medium is then given by
\begin{equation}\label{eq:surfr}
\Sigma_R=2\int_{H_D\simeq 0}^\infty\rho_R(z)dz = \SigNG
\mathcal{F}(s_0),
\end{equation}
where
\begin{equation}\label{eq:SigNG}
\SigNG \equiv (2\pi)^{1/2}h_g \rho_R(0)
=\frac{P_R(0)}{\sqrt{ 2G\rho_*c_R^2}}
\end{equation}
is the surface density without gas self-gravity,
\begin{equation}\label{eq:reduc}
\mathcal{F}(s_0)\equiv\exp(s_0){\rm erfc}(s_0^{1/2})
\end{equation}
is the reduction factor, and
\begin{equation}
s_0\equiv\frac{\pi G\Sigma_D^2}{2c_R^2 \rho_*},
\end{equation}
measures the strength of gravity due to the dense gaseous slab relative
to the external vertical gravity (see, e.g., \citealt{kos02}).
Note that the results of
\S\ref{sec:vequil} suggest that when two-phase equilibria are
established, the interface between dense and rarefied media has
a constant pressure $P_{\rm trans}$, so that we may take
$\rho_R(0)=P_R(0)/c_R^2=P_{\rm trans}/c_R^2$ since the dense medium has a very
small scale height (e.g., \citealt{pio07}), such that
\begin{equation}
\Sigma_R=\frac{  P_{\rm trans} \mathcal{F}(s_0)}{  \sqrt{2 G \rho_* c_R^2}}.
\label{eq:SigR}
\end{equation}
The scale height of the
rarefied medium is given by
\begin{equation}
H_R^2=\frac{\int_{-\infty}^\infty
\rho_R(z)z^2dz}{\int_{-\infty}^\infty\rho_R(z)dz}=
h_g^2\left[(1+2s_0)-\sqrt{\frac{s_0}{\pi}}\frac{2}{\mathcal{F}(s_0)}\right],
\end{equation}
where equation (\ref{eq:rhor}) is used.

The condition of mass conservation requires
\begin{equation}\label{eq:massr}
\Sigma_0=\Sigma_D + \Sigma_R
\end{equation}
so that
\begin{equation}
s_0=\frac{\pi G (\Sigma_0-\Sigma_R)^2  }{2 c_R^2 \rho_*}.
\label{eq:s0}
\end{equation}
For self-gravitating cases, we fix $\rho_*=0.056\Msun\pc^{-3}$
and $c_R=7\kms$, and solve equations (\ref{eq:SigR}) and (\ref{eq:s0})
iteratively to find $\Sigma_R$ and
$\Sigma_D=\Sigma_0-\Sigma_R$ as functions of $\Sigma_0$.
The resulting values of $f_R=\Sigma_R/\Sigma_0$ and $H_R$ are plotted in
Figure \ref{fig:mf0} as dashed lines. For non-self-gravitating
cases, $\Sigma_R= \SigNG
=P_R(0)/(2G\rho_*c_R^2)^{1/2}$
and $H_R=h_g$ corresponding to $s_0=0$,
plotted as solid lines in Figure~\ref{fig:mf0}.

Note that since $\Sigma_D<\Sigma_0$, $s_0\ll 1$ if $\pi
G\Sigma_0^2/(2c_R^2 \rho_*)\ll 1$; for our models with $\rho_*=0.056
\Msun \pc^{-3}$ and $\Sigma_0< 10 \Msun \pc^{-2}$, $\pi
G\Sigma_0^2/(2c_R^2 \rho_*)< 0.2$.  When $s_0\ll 1$, ${\cal F}(s_0)
\approx 1$, so that $H_R \approx h_g \rightarrow 128 \pc$ and
$\Sigma_R\rightarrow 2.8 \Surf$ for $\rho_*= 0.056 \Msun \pc^{-3}$
and $P_{\rm trans}/\kb = 2100 \Kel \pcc$.

\end{document}